\documentclass[12pt,aps,prd,preprint,onecolumn,nofootinbib,superscriptaddress]{revtex4-2}

\usepackage{amsmath,amssymb,mathtools}
\usepackage{mathrsfs}
\allowdisplaybreaks
\usepackage{slashed}
\usepackage{physics}
\usepackage{enumitem}
\usepackage{cancel}
\usepackage{graphicx}
   \usepackage{xcolor}

\usepackage{url}
 \usepackage{comment}

\usepackage[normalem]{ulem}

\newcommand{\rmax}{r_\text{max}}
\newcommand{\vmax}{V_\text{max}}
\newcommand{\gnewton}{G_\mathrm{N}}

\newcommand{\Msun}{M_\odot}

\usepackage{tikz}
\usetikzlibrary{decorations.pathmorphing}

\usepackage{microtype}
\usepackage{url}

\usepackage[colorlinks=true,allcolors=black]{hyperref}

\usepackage[most]{tcolorbox}

\newtcolorbox{computation}[1][]{%
    enhanced,
    breakable,
    colback=gray!5,
    colframe=gray!50,
    fonttitle=\bfseries,
    title={Detailed Computation},
    #1
}
\makeatletter
\renewcommand{\p@subsection}{}
\renewcommand{\p@subsubsection}{}
\makeatother

\begin{document}
%=========================================================

%\title{When Black Holes Can Wear Pants}
\title{{When Black Holes Admit Pair-of-pants Topologies}}

\author{Giacomo Cacciapaglia}
\affiliation{Laboratoire de Physique Th\'eorique et Hautes \'Energies (LPTHE), UMR 7589,\\ Sorbonne Universit\'e \& CNRS, 4 place Jussieu, 75252 Paris Cedex 05, France}
\affiliation{\mbox{$\hbar$QTC, Quantum Theory Center, \& Danish Institute for Advanced Study (Danish IAS)}, University of Southern Denmark, Campusvej 55, DK-5230 Odense M, Denmark}

\author{Manuel Del Piano}
\affiliation{\mbox{$\hbar$QTC, Quantum Theory Center, \& Danish Institute for Advanced Study (Danish IAS)}, University of Southern Denmark, Campusvej 55, DK-5230 Odense M, Denmark}

\author{Francesco Sannino}
\affiliation{\mbox{$\hbar$QTC, Quantum Theory Center, \& Danish Institute for Advanced Study (Danish IAS)}, University of Southern Denmark, Campusvej 55, DK-5230 Odense M, Denmark}
\affiliation{\mbox{Dipt. di Fisica ``E. Pancini'', Universit\`a di Napoli Federico II, Via Cintia, 80126 Napoli, Italy}}

\author{Vania Vellucci}
\affiliation{\mbox{$\hbar$QTC, Quantum Theory Center, \& Danish Institute for Advanced Study (Danish IAS)}, University of Southern Denmark, Campusvej 55, DK-5230 Odense M, Denmark}
%\date{}

\begin{abstract}
We investigate the conditions under which black hole fragmentation, the splitting of a black hole into {two or} multiple smaller ones, may occur {via an unspecified dynamics, which would have the topology of a pair-of-pants. We employ kinematical and entropical consideration on time-sliced quasi-static black hole configurations to check under what conditions the pants dynamics is permitted.}  %\sout{The simplest realization is that of a single black hole horizon splitting into two, giving rise to the eponymous pants topology.} 
In classical general relativity, the Bekenstein–Hawking area law forbids such processes for Schwarzschild black holes. For spinning Kerr black holes, purely kinematic analyses impose constraints that prevent fragmentation, even in regimes where entropy considerations might allow it, except possibly in near-extremal cases.
We then hunt for scenarios where black holes can wear pants: from the well-known Gregory--Laflamme instability in higher dimensions, to the potential effect of superradiant instabilities in non-axisymmetric radiation trapping, to finally gravitational models that modify the relations between entropy and/or horizon radius and the black hole mass in four dimensions. In all such cases, emission of small fragments can be entropically favored,  however its occurrence still depends on the kinematic configuration of the initial state. 
Our analysis clarifies the theoretical landscape where black holes may fragment, which is particularly relevant for primordial black holes and catastrophic events such as black hole mergers.
 \end{abstract}

\maketitle

\newpage 
\setcounter{tocdepth}{2}
\tableofcontents
\bigskip

\newpage

\section{Introduction}

\begin{figure}
    \centering
    \includegraphics[width=0.4\linewidth]{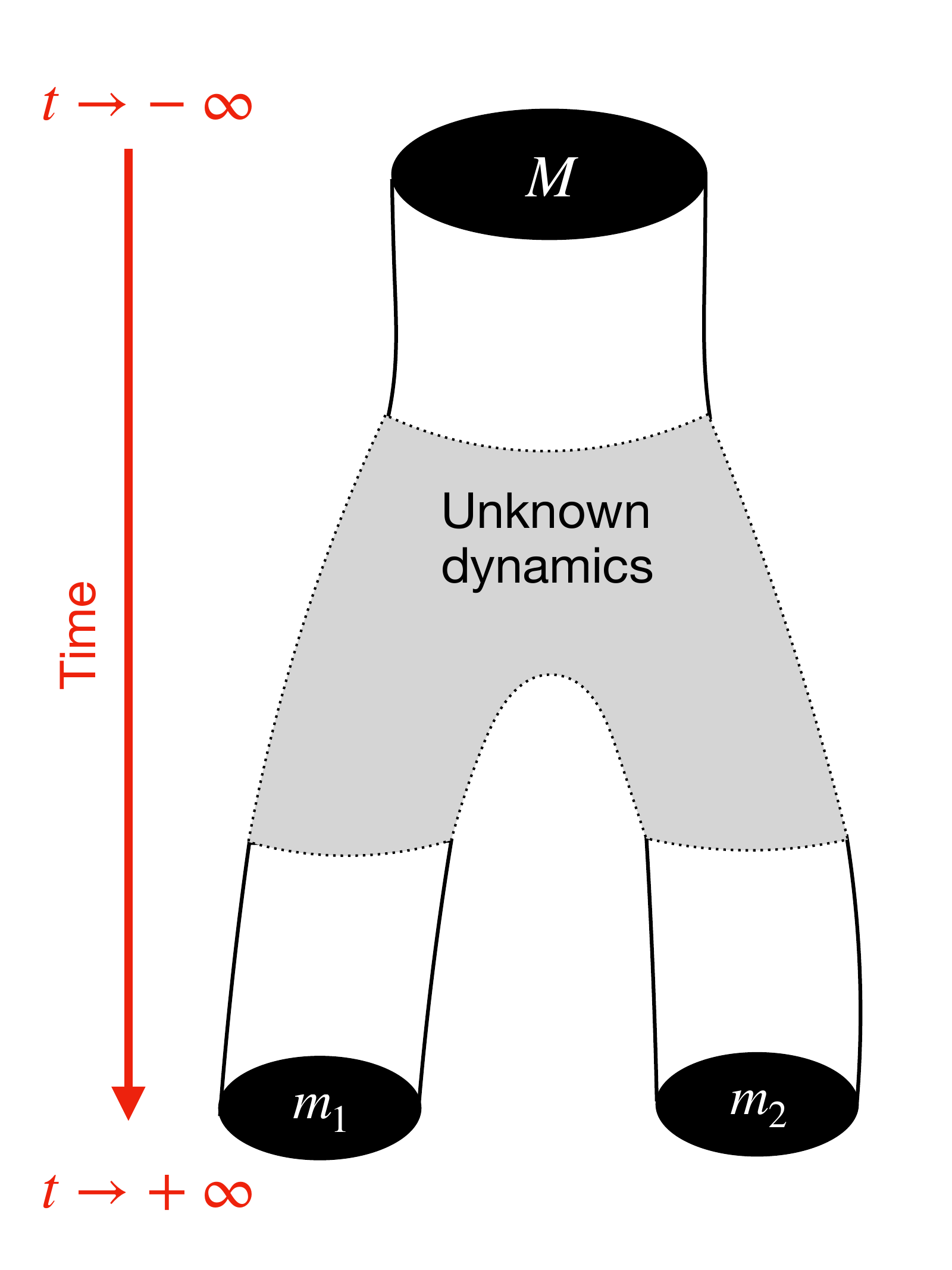}
    \caption{{Diagrammatic representation of an asymptotic spacetime containing a single black hole of mass $M$ evolving into a configuration of two black holes with masses $m_1$ and $m_2$ via an unknown dynamics. If a continuous horizon connects the two states, it has the topology of a \textit{pair of pants}.} }%\sout{Simplest diagram representing black hole pants: a single black hole of mass $M$ splitting into two fragments of masses $m_1$ and $m_2$.}}
    \label{fig:BHpants}
\end{figure}

Answering the question whether a black hole can \textit{``wear pants''}, that is, whether a single horizon can evolve into a configuration that effectively resembles multiple black holes, is a  probe of the interplay between dynamics, thermodynamics, and global structure in gravity. The simplest illustration of this process would be a single black hole horizon evolving into two separate ones, see Fig.~\ref{fig:BHpants}, which showcases the eponymous topology. In four-dimensional classical general relativity (GR), stationary black holes appear remarkably rigid: \textit{Hawking's area theorem}~\cite{Hawking:1971tu} constrains any process that tends to reduce the total area of the horizon, as it would lead to a decrease of the total entropy stored in the black holes, {a statement that has recently received direct observational support through the gravitational signal of binary mergers ~\cite{LIGOScientific:2025rid,Tang:2025jyj}}. As such, a single Schwarzschild black hole is always entropically favored over two black holes of the same total mass. Furthermore, the linear stability of spinning Kerr black holes under small perturbations has been well established ~\cite{Whiting:1988vc,Giorgi:2022omp}. At face value, these facts suggest that spontaneous fragmentation of an isolated black hole into smaller constituents is forbidden. Yet this conclusion is contingent on assumptions, notably suitable energy conditions and a form of cosmic censorship, and it can fail in settings such as higher-dimensional spacetimes, higher-curvature theories, or semiclassical regimes. Understanding precisely what is {forbidden}, what is {merely unlikely}, and what {becomes possible} once the assumptions are relaxed is the central goal of this work.

{Throughout this work, by \textit{black hole} we mean the region enclosed by the outermost marginally trapped surface (apparent horizon, understood as slice-dependent); this is the object whose merger or bifurcation we analyze in a given foliation. We reserve the term \textit{event horizon} for the strictly global, teleological boundary $\partial J^{-}(\mathscr{I}^{+})$, which in GR and stationary configurations coincide with the apparent horizon and which enters only in the topological \textit{pants} statement and in Penrose's non-bifurcation theorem. } This theorem is related to the global properties of event horizons. It states that, in a strongly asymptotically predictable spacetime, horizon generators cannot have future endpoints~\cite{Wald:1984rg}; heuristically, event horizons do not split in such spacetimes and black holes do not wear pants. The theorem does not exclude every notion of ``fragmentation'', instead it clarifies what must give way: a genuine horizon bifurcation in a classical setting would require a breakdown of strong asymptotic predictability, which is typically associated to the appearance of naked singularities. Note, however, that in a black hole scattering not every multi-object outcome requires horizon splitting: for instance, a $2\to 2$ process, where two black holes interact and separate into two new black holes of different parameters~{\cite{Sperhake:2008ga}}, may not involve a bifurcation of a single connected horizon, provided no common trapped region forms during the interaction.

If fragmentation were possible under certain conditions, the observational consequences could be significant. It would provide an alternative formation channel for black holes with masses outside the ranges accessible through standard astrophysical processes, potentially producing  black hole {\it morsels} small enough to exhibit observable Hawking evaporation on experimentally accessible timescales~\cite{Cacciapaglia:2024wtp,Acharyya:2025yzx}. Conversely, the observed stability of black holes across a wide range of masses and spins places stringent constraints on any fragmentation scenario.

In our analysis, we will be mainly concerned with entropy considerations, i.e. the fact that states with larger entropy are statistically favored, including kinematics when possible, to study black hole fragmentation within various contexts. {Henceforth, we will consider time-slices of space-time containing approximately static horizons, which we identify as black holes, and compare the entropy and kinematical properties of slices with one or two such objects. We will not be concerned with the dynamics connecting the two states in a fragmentation process (i.e. the proper pants topology in Fig.~\ref{fig:BHpants}), which requires non-equilibrium space-time geometry.}
Within GR, the basic entropic obstruction to fragmentation is immediate for non-rotating black holes with Schwarzschild metric, for which the entropy is proportional to the mass squared, i.e. $S(M) \propto GM^{2}$, where $G = 1/M_\text{P}^2$ is Newton's constant in terms of the Planck mass $M_\text{P}$. Being a convex function of the mass, if one imagines a putative splitting process, $1 \to 2$, with $M = m_{1} + m_{2}$ see Fig.~\ref{fig:BHpants}, then $S(m_{1}) + S(m_{2}) \leq S(M)$, so fragmentation would violate the second law of thermodynamics. The rotating case is more subtle: for a Kerr black hole of given mass, the larger the spin the smaller the entropy. Hence, in principle one can envision a configuration where some, if not all, of the original angular momentum could be converted into orbital angular momentum of the fragments, allowing the total entropy of the final configuration to compete with the initial one. However, any such channel must also satisfy energy and angular momentum conservation together with the requirement that the fragments actually separate rather than promptly re-merge. We make this tension explicit and we delineate, in the cleanest possible way, the combined entropic and kinematic obstructions to true fragmentation within classical four-dimensional GR.

Higher-dimensional GR dramatically changes the situation and provides a concrete arena where fragmentation-like evolution is not only entropically plausible, but dynamically realized. Extended horizons, such as black strings, suffer from the Gregory--Laflamme instability~\cite{Gregory:1993vy}, and ultra-spinning Myers--Perry black holes~\cite{Myers:1986un} develop instabilities that can be understood as approaching the physics of extended black branes~\cite{Emparan:2003sy,Emparan:2014jca}. In these cases, the ``\textit{black hole pants}'' picture becomes literal: the horizon evolves toward highly inhomogeneous configurations, and the endpoint can involve pinch-off behavior and violations of cosmic censorship in classical evolution~\cite{Lehner:2010pn}. These examples teach us an essential lesson: entropic reasoning is a powerful diagnostic, but the actual fate of an instability is ultimately dynamical and may involve singular behavior once the classical assumptions are relaxed. Motivated by these lessons, we broaden the discussion toward two additional directions. 

First, we consider {\it effective fragmentation} mechanisms in which a Kerr black hole may come to behave {\it as if} it had developed companions without a literal horizon bifurcation. A representative possibility is provided by superradiant amplification~\cite{Press:1972zz} in the presence of asymmetric trapping, which may occur due to plasma-induced effective masses for photons or additional light degrees of freedom in a non-uniform accretion disk~\cite{Conlon:2017hhi}.

Second, we examine how fragmentation criteria are altered beyond GR, where the mass-dependence of the horizon radius and/or entropy is affected. As an example, we consider higher-derivative gravity theories~\cite{Bueno:2016lrh,Bueno:2017qce}, where both the radius and the entropy functional (via Wald entropy) differ from the GR ones. This can in principle reverse the entropic balance for sufficiently asymmetric mass splits, when one fragment is smaller than a characteristic scale, hence making fragmentation thermodynamically favored even in four dimensions. Taking into account current bounds, fragmentation may be entropically favored even for astrophysical black holes. We stress, nevertheless, that an entropic preference is not a proof of dynamical instability; it is instead a guide to where a genuine instability might exist, to be confirmed by nonlinear dynamical analyses. The same results could be obtained in more general modifications of the entropy relation to the black hole mass.

Along the same direction, we find that non-trivial horizon topologies induced by external fields may also enable fragmentation. Recent work has shown that bumpy horizon geometries can be supported by physically realistic matter configurations, such as superfluid pion condensates developing vortex structures~\cite{Canfora:2026col}. The topological charges associated with these vortices modify both the Arnowitt-Deser-Misner (ADM) mass and the entropy, opening a window for entropically allowed fragmentation even within four-dimensional GR, provided the normalized topological charge is sufficiently large.

The manuscript is organized as follows. In Sec.~\ref{sec:GR} we present the classical GR constraints on black-hole fragmentation, combining the area theorem with kinematic considerations for both Schwarzschild and Kerr black holes. %; we show that while entropy considerations alone might suggest channels for Kerr fragmentation, separating fragments that persist to infinity is obstructed by energy and angular-momentum constraints in the regime of interest. In Sec.~\ref{sec:higherdim} 
Then, we review the higher-dimensional scenarios of black strings and ultra-spinning Myers--Perry black holes, where fragmentation-type instabilities arise and have been confirmed numerically. In Sec.~\ref{sec:superradiance} we discuss to what extent  superradiance in nontrivial environments could lead to effective fragmentation. In Sec.~\ref{sec:modified} we analyze how higher-curvature corrections and more general deformations of the entropy functional modify the thermodynamic criteria for fragmentation, and we identify the parameter regimes where such effects become relevant. We also discuss how bumpy horizons supported by topological defects in superfluid pion configurations provide an additional, and qualitatively distinct, fragmentation channel. We conclude in Sec.~\ref{sec:conclusions} with a synthesis of the conditions under which black holes may, in one sense or another, {\it wear pants}.

\section{General Relativity constraints on Black Hole pants, a brief summary}
\label{sec:GR}

We will first critically analyze why common lore says that black hole fragmentation is forbidden in GR by the second law of black hole thermodynamics. We will see that this is no more true in more than four dimensions. Even for a four-dimensional rotating black hole, simple kinematic considerations allow the formation of fragments, at least in theory. {We emphasize at the outset that an entropy increase is a \textit{necessary but not sufficient} condition for fragmentation: it is a universal filter that any second-law-respecting channel must pass, but by itself it certifies neither a dynamical instability nor an accessible path to a multi-horizon endpoint. %(much as a glass has higher entropy once shattered, yet does not spontaneously shatter). 
Wherever possible we therefore supplement the entropic test with kinematic conditions or with independent dynamical evidence, and we flag explicitly those channels for which only an entropic preference has been established. We also stress the scope of our claims: throughout we track \textit{horizons}, not the central singularity. We do not attempt to bisect a curvature singularity or its distributional source; a literal event-horizon bifurcation, when it occurs (as in the Gregory--Laflamme evolution), proceeds through a pinch of diverging curvature, i.e.\ a classical naked singularity, consistent with the breakdown of strong asymptotic predictability.}

{In what follows, unless stated otherwise, $M$ denotes the total ADM mass of the asymptotically flat initial spacetime, while $m_i$, $\mu$ denote the rest masses, equivalently the ADM masses in their individual asymptotic rest frames, of well-separated final black holes. Exact conservation allows for radiation and binding/kinetic energy,
\begin{equation}
 M_{\rm ADM}=E_1^{\infty}+E_2^{\infty}+E_{\rm rad},\qquad
 J_{\rm ADM}=J_1+J_2+L_{\rm orb}+J_{\rm rad}.
 \label{eq:asymptotic-conservation}
\end{equation}
Our entropy-only estimates adopt the idealized zero-radiation limit, which is explicitly stated when used. In the Kerr point-particle model below, the kinetic and gravitational-potential costs are retained through the specific energy $E$ and effective potential $V(r)$, so that $M=m+\mu E$ and $J=\mu L$ in the simplified nonspinning-fragment configuration.}

\subsection{Schwarzschild black holes}
Assuming that the entropy is given by the Bekenstein--Hawking formula~\cite{Bekenstein:1972tm,Bekenstein:1973ur,Hawking:1974rv}, the entropy of a Schwarzschild black hole of mass $M$ scales as
\begin{equation}
    S_M \propto G M^2\ ,
\end{equation}
which is a convex function of the mass. Supposing that a Schwarzschild black hole of initial mass $M$ fragments into two smaller black holes of masses $m_1$ and $m_2$, where conservation of mass requires $M=m_1+m_2$, we have
\begin{equation}
    S_{m_1}+S_{m_2}\propto G (m_1^2+m_2^2)< S_M \propto G (m_1+m_2)^2 \ .
\end{equation}
Hence, the configuration of two separate black holes is always entropically disfavored with respect to a single black hole of equal mass, rendering the fragmentation unlikely.

\subsection{Kerr black holes}
It is well known that rotating Kerr black holes are stable, at least at linear level \cite{Whiting:1988vc} (for recent progress about the full non-linear stability see \cite{Giorgi:2022omp}). However, disfavoring the possibility of fragmentation from entropic arguments alone is less trivial than it was for static black holes.
Let us, again,  assume the Bekenstein--Hawking entropy formula for a rotating Kerr black hole with mass $M$ and angular momentum $J$. Its entropy scales as 
\begin{equation} \label{eq:Kerrenropy}
    G S \propto (GM+\sqrt{G^2 M^2-a^2})^2 +a^2 \qq{with} a \coloneqq J/M \ ,
\end{equation}
which is a monotonically decreasing function of the rotation parameter $a\in[0,GM]$ (let us consider, without loss of generality, positive $a$, i.e. counter-clockwise rotations). Hence, slowly rotating black hole states are thermodynamically favored with respect to highly rotating ones.
As a consequence, in principle, a highly rotating black hole might fragment without decreasing the entropy if part of its intrinsic angular momentum (spin) becomes angular momentum of the new binary system. However, this implies that part of the initial energy (i.e., the initial mass) is transformed into kinetic energy in order to impart a velocity to the fragment. This effect leads to a reduction of the entropy of the final state.
We will show how the fragmentation process, while entropically possible, is  prevented by kinematic and conservation laws, except for nearly extremal initial black holes.

\begin{figure}
    \centering
    \includegraphics[width=\linewidth]{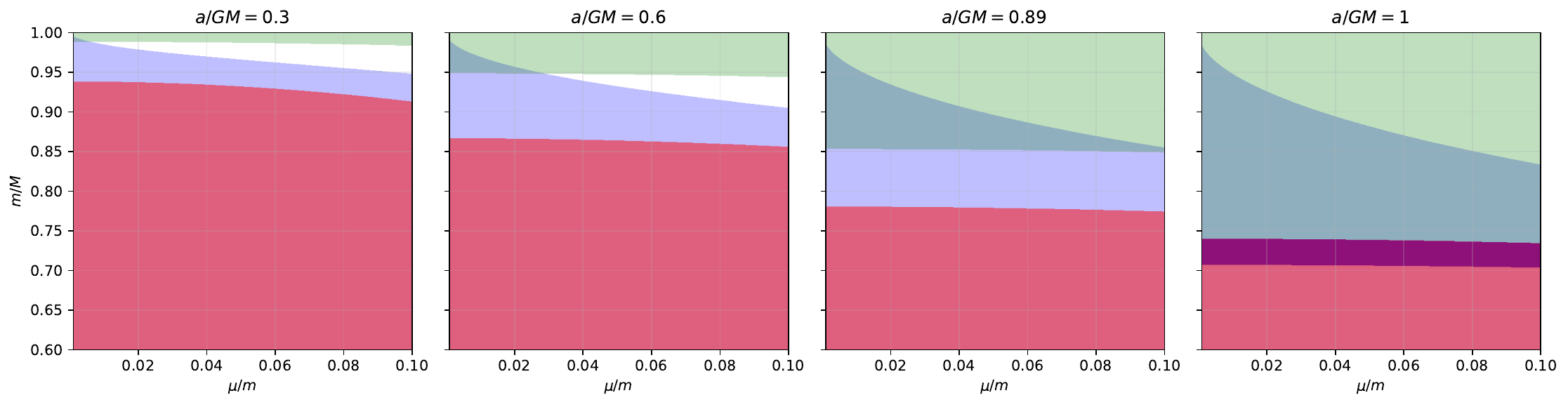}
    \caption{Regions in the parameter space of the two non-rotating fragments for four representative values of the initial Kerr rotation parameter $a$, normalized to the initial mass. As the analysis relies on the point-particle approximation, the mass of the smaller fragment is expressed relative to the larger one as $\mu/m$. The shaded areas correspond to the domains where the three fragmentation conditions are individually satisfied: red for the escape condition in Eq.~\eqref{condition on energy}, blue for the velocity condition in Eq.~\eqref{condition on velocity} and green for the entropy condition in Eq.~\eqref{condition on entropy}. The purple region in the rightmost panel highlights their simultaneous intersection.}
    \label{fig:Kerr-conditions}
\end{figure}

We consider, for simplicity, a highly rotating (and thus low entropy) Kerr black hole fragmenting into two non-rotating black holes with masses $m$ and $\mu$: we work in the limit where one of the two fragments is much lighter than the other ($\mu \ll m$). In this way, we can approximate the smaller fragment as a test particle moving in the background of the bigger one, given by the usual Schwarzschild metric:
\begin{equation}
    \dd s^2= -\left(1- \frac{2 G m}{r} \right) \dd t^2 + \left(1-\frac{2 G m}{r}\right)^{-1}\dd r^2 + r^2 \dd \Omega_2^2 \ .
\end{equation}
We assume that the two fragments with radius $2Gm$ and $2G\mu$, respectively, are produced one near the other, thus the initial radial distance between their centers is $2G(m+ \mu)$. We call the \textit{coordinate velocity} imparted to $\mu$ during the fragmentation $v_r\coloneqq \dd r / \dd t$. After the fragmentation, the smaller fragment moves along a geodesic on the background given by the bigger fragment, with initial conditions: $r|_{t=0}\equiv r_0=2 G(m + \mu)$ and $\left.\dd r/\dd t\right|_{t=0}=v_r$. The Lagrangian per unit mass for a point particle, alias the smaller fragment, reads
\begin{equation}
    \mathcal{L}= \frac{1}{2}g_{\mu \nu}u^\mu u^\nu \ ,
\end{equation}
where $u^{\alpha}\coloneqq\dv{x^{\alpha}}{\tau}=(u^t,v_r \, u^t, 0, u^\phi) $ is the four-velocity of the point-like fragment of mass $\mu$ and $\tau$ is an affine parameter along the geodesic.
Given the spherical symmetry of the background we can study the motion on the equatorial plane ($\theta=\pi/2$) and we have two conserved quantities along the geodesic:
\begin{equation}
    E= u^t \left(1-\frac{2 G m}{r}\right) \qq{and} L=  u^{\phi} r^2 \ ,
\end{equation}
which respectively represent the energy and the angular momentum per unit mass of the smaller fragment.
Hence, the four-velocity can be written as
\begin{equation}
    u^{\alpha} = \left(\frac{E}{1-\frac{2 G m}{r}},\frac{E \, v_r }{1-\frac{2 G m}{r}}, 0,\frac{L}{r^2}\right) \ ,
\end{equation}
and, from the norm of the four-velocity $u_\alpha u^\alpha = -1$, we can solve for the energy $E$
\begin{equation}\label{m2 energy}
    E^2 =  \left( 1- \frac{v_r^2}{\left(1-\frac{2Gm}{r}\right)^2}\right)^{-1}V(r) \ ,
\end{equation}
where $V(r)$ is the effective gravitational potential along timelike geodesics, which reads
\begin{equation}
    V(r)= \left(1-\frac{2G m}{r} \right)\left(1+\frac{L^2}{r^2} \right) \ .
\end{equation}
The position of the local maximum of the potential $\rmax$ is
\begin{equation}
    \rmax = \frac{L}{2G m}\left( L - \sqrt{L^2 - 12 G^2 m^2}\right) \ ,
\end{equation}
which exists if $L\geq 2 \sqrt{3} \, G m$, otherwise the potential monotonically increases from $V(2G m)=0$, at the horizon, to $V(r\to \infty)\to 1$. From the conservation of the angular momentum and energy, we can relate the parameters to the mass $M$ and spin $a$ of the initial Kerr black hole
\begin{equation}\label{conserved L and E}
    L=\frac{a M}{\mu} \qq{and} E=\frac{M-m}{\mu} \ ,
\end{equation}
so the condition of existence for the maximum of the potential reads
\begin{equation}
    aM \geq 2 \sqrt{3} G \, m \, \mu \ , \label{max}
\end{equation}
and the value of the potential at the local maximum is
\begin{equation}\label{vmax}
    \vmax = \frac{2\left(1-\frac{4 G^2 m^2 \mu^2}{(aM)^2}- \sqrt{1-\frac{12 G^2 m^2 \mu^2}{(aM)^2}}\right)^2}{\left(1-\sqrt{1-\frac{12 G^2 m^2 \mu^2}{(aM)^2}}\right)^3} \ .
\end{equation}
In order for the smaller fragment to escape the gravitational pull of the bigger fragment, we must impose that $E^2 > \vmax$.  Using Eqs~\eqref{conserved L and E} and \eqref{vmax}, we can, therefore, express the \textit{escape condition} as follows:
\begin{equation}\label{condition on energy}
    \frac{M-m}{\sqrt{2}\, \mu} - \frac{1 - \frac{4 G^2 m^2 \mu^2}{(aM)^2} - \sqrt{1-\frac{12 G^2m^2 \mu^2}{(aM)^2}}}{\left(1-\sqrt{1-\frac{12 G^2m^2 \mu^2}{(aM)^2}}\right)^{3/2}} > 0 \  ,
\end{equation}
or $E^2>1$ if the initial spin does not satisfy Eq.~\eqref{max}.
In addition, one has to check that the radial velocity of the escaping fragment is positive and well defined, which happens when the following \textit{velocity condition} is satisfied:
\begin{equation}\label{condition on velocity}
    0<v_r < 1-\frac{2 G m}{r_0}  \implies 0<\frac{V(r_0)}{E^2}=\frac{\mu((a  M)^2+4 G^2 \mu^2( m + \mu)^2)}{4 G^2(M- m)^2(m + \mu)^3}<1 \ .
\end{equation}
Finally, one must take into account the increase in entropy following the fragmentation, which leads to the following \textit{entropy condition}:
\begin{equation}\label{condition on entropy}
   \frac{ 4 G^2( \mu ^2 +  m ^2)}{(G M+\sqrt{G^2M^2-a^2})^2 + a^2 } > 1 \ .
\end{equation}
The restrictions contained in Eqs.~\eqref{condition on energy},~\eqref{condition on velocity} and ~\eqref{condition on entropy} are typically incompatible with one another, in particular the entropy requires the main fragment to be as large as possible, while the escape condition prefers small masses $m$. This is illustrated in  Fig.~\ref{fig:Kerr-conditions} for different values of the rotation parameter $a$. In particular, the parameter space selected by the conditions in Eq.~\eqref{condition on velocity} and Eq.~\eqref{condition on entropy} can always intersect, implying that production of a fragment in the vicinity of the horizon may occur, however it will be reabsorbed by the horizon after a small finite time due to the failing of the escape condition. We only find an intersection of condition \eqref{condition on energy} with the other two when the initial Kerr black hole is (nearly) extremal, indicating that such fragment must have a large angular momentum to be capable of escaping to infinity. {This near-extremal restriction has a transparent thermodynamic origin. The Kerr entropy in Eq.~\eqref{eq:Kerrenropy}, $GS\propto(GM+\sqrt{G^2M^2-a^2})^2+a^2$, has a spin-gradient $\partial S/\partial a\propto -\,GMa/\sqrt{G^2M^2-a^2}$ that diverges as $a\to GM$, where $T\to0$. Hence, near extremality, the entropy becomes extremely sensitive to despinning: a small transfer of spin into orbital angular momentum produces a large entropy change, which can offset the finite entropy cost of the kinetic energy required for the fragment to escape. Away from extremality  the kinematic cost forbids escape.} 

In Figure~\ref{fig:Kerr-conditions} there always exists an intersection between the entropy and the velocity conditions, i.e. the overlap of the green and blue areas. In this region, the fragment can be formed but cannot escape, hence it is reabsorbed after a finite time. It is natural to wonder if the reabsorption time is longer or shorter than the evaporation time, potentially leading to observable effects or the absence thereof. To answer this question in simple terms, we put forward an ansatz on the kinematics involved in the fragmentation: given the range of the admissible parameters, let us assume that the mass loss due to the energy transferred to the smaller fragment scales as follows{, subject to the modeling requirement that the energy drained from the parent, $M-m$, be subdominant as $\mu\to0$, so that the background is undisturbed in the point-particle limit,}
\begin{equation}\label{energy-transfer-alpha}
{\frac{M-m}{\mu} = f(\mu, m) \qq{such that}  \lim_{\mu \to 0} \mu \,f(\mu , m) = 0  \implies }
    M-m \sim  \alpha \left( \frac{\mu}{m}\right)^p\mu 
\end{equation}
for some real constants $\alpha > 0$ and $p>-1$. This allows us to compute the flying time as 
\begin{align}
    \tau &\approx 2\int_{r_0}^{r_\ast} \frac{\dd r}{\sqrt{E^2 - (1-2Gm/r)(L^2/r^2+1)}} \\
    & = \frac{2}{\alpha (\mu/m)^p} \int_{r_0}^{r_\ast} \frac{\dd r}{\sqrt{\mathcal{E}(r)}} \ , \label{proper time integral}
\end{align}
where the integral runs from the initial position $r_0= 2G(m+\mu)$ to the turning point $r_\ast$. The latter is defined by the zero of the function
\begin{equation}
    \mathcal{E}(r_\ast) \coloneqq 1 - \left(1-\frac{2Gm}{r_\ast}\right)\left( \frac{b^2}{r_\ast^2}+\frac{m^{2p}}{\alpha^2 \mu^{2p}}\right)= 0 \qq{with} b = \frac{L}{E}= a\left(1 + \frac{1}{\alpha}\left(\frac{m}{\mu}\right)^{p+1} \right)  \ ,
\end{equation}
where we used Eq.~\eqref{conserved L and E} to express $L$ and $E$ in terms of $\mu$ and $m$.
The integral in Eq.~\eqref{proper time integral} is of elliptic type and is usually computed by means of Jacobi's elliptic functions. However, here we follow a perturbative approach using the mass of the flying fragment as expansion parameter.
Expanding in $\mu/m \ll 1$, the relevant contribution in the second bracket can be estimated around the horizon $r= 2Gm + \delta$, with $\delta \ll 2G m$, 
\begin{equation}
    b^2/r^2 = \frac{a^2(1+m^{p+1}/\alpha \mu^{p+1})^2}{(2Gm+\delta)^2} \approx \frac{a^2 m^{2p}}{4G^2\alpha^2 \mu^{2p+2}} \ ,
\end{equation}
so that we can compare with the constant term in the brackets
\begin{equation}
    \frac{b^2 /r^2}{m^{2p}/\alpha^2 \mu^{2p}} \approx \frac{a^2}{4G^2m^2}\left( \frac{m}{\mu}\right)^{2} \gg 1 \ ,
\end{equation}
hence, the relevant contribution comes from the centrifugal barrier (as in the eikonal limit).
Then, the equation for the turning point reads
\begin{equation}
    \left(1-\frac{2Gm}{r_\ast}\right) \frac{b^2}{r_\ast^2} = 1 \ .
\end{equation}
Going near the horizon $r_\ast = 2Gm + \delta_\ast$, we have at leading order:
\begin{equation}
    \frac{\delta_\ast \, b^2}{8 G^3m^3} = 1 \implies \delta_\ast \approx \frac{8 \alpha^2 G^3m^3}{a^2}\left( \frac{\mu}{m}\right)^{2p+2} \ ,
\end{equation}
thus we can compute the interval
\begin{equation}\label{turning-point-position}
    \frac{r_\ast - r_0}{2G m} =  \frac{4 \alpha^2 G^2m^2}{a^2} \left( \frac{\mu}{m}\right)^{2p+2} - \frac{\mu}{m} =  \frac{\mu}{m}\left(\frac{4 \alpha^2 G^2m^2}{a^2}\left(\frac{\mu}{m}\right)^{2p+1}-1 \right) \ .
\end{equation}
{For $-1<p \leq -1/2$, Eq.~\eqref{turning-point-position} can be positive, so that} the turning point $r_\ast$ can be taken arbitrarily larger than $r_0$ (as long as energy conservation allows it). On the contrary, for $p>-1/2$, the interval cannot be made positive, meaning that the morsel starts from $r_0$ with an energy below the potential, so that its motion is forbidden. Let us consider the borderline case $p=-1/2$, for which, using the relation in Eq.~\eqref{energy-transfer-alpha}, we have
\begin{equation}
   \alpha > \frac{a}{2 G m} \implies \frac{a}{2GM}\sqrt{\frac{\mu}{m}}+\frac{m}{M} < 1\ .
\end{equation}
%{This positivity requirement is not in tension with the centrifugal dominance found above: the ratio estimated near the horizon (centrifugal versus constant term) is large because of the angular-momentum factor $b=L/E\propto(m/\mu)$ and is independent of $p$, whereas the sign of the interval is controlled by the amplitude $\alpha$ through the combination $4\alpha^2G^2m^2/a^2\,(\mu/m)^{2p+1}$. At the borderline $p=-1/2$ the exponent $2p+1$ vanishes, the $\mu/m$ suppression cancels, and the condition reduces to the $\mathcal{O}(1)$ requirement $\alpha>a/(2Gm)$; no large negative $p$ is needed.}
Since the integrand in Eq.~\eqref{proper time integral} is monotonically increasing in a parametrically small interval controlled by $\mu$ and $\alpha$, we can approximate it by expanding the function
\begin{equation}
    \mathcal{F}(r) \coloneqq 1 - \left(1-\frac{2Gm}{r}\right) \frac{b^2}{r^2} \ ,
\end{equation}
around the turning point $r_\ast$, yielding
\begin{align}
    \tau &\approx \frac{2}{\alpha}\sqrt{\frac{\mu}{m}}\int_{r_0}^{r_\ast}\frac{\dd r}{\sqrt{-\mathcal{F}^\prime(r_\ast) (r_\ast-r)}}=\frac{4}{\alpha\sqrt{-\mathcal{F}^\prime(r_\ast)}}\sqrt{\frac{\mu}{m}}\sqrt{r_\ast-r_0}  = \\
    &=\frac{16 G^2 \mu}{a}\left(1 + \frac{4 \alpha^2 G^2 m \mu}{a^2} \right)^2\sqrt{ \frac{m \mu \left(\frac{4\alpha^2 G^2m^2}{a^2}-1 \right)}{1 - \frac{8 \alpha^2 G^2 m \mu}{a^2}}} \\
    &\approx \frac{32G^3 (M-m)\, m\,\mu}{a^2}\left(1+ \frac{4 G^2 (M-m)^2}{a^2} \right)^2 \left(1-\frac{8G^2 (M-m)^2}{a^2} \right)^{-1/2} \ , %+ \order{\left( \frac{\mu}{m} \right)^{5/2}}\ ,
\end{align}
where in the last line we used the relation in Eq.~\eqref{energy-transfer-alpha} for $p=-1/2$ to fix $\alpha$.
We can compare this time scale with the evaporation time $t_\mathrm{evap} = \gamma G^2 \mu^3$, where $\gamma = 5120 \pi$ for a single massless bosonic degree of freedom and no gray-body corrections~\cite{Hawking:1975vcx}, leading to the \textit{evaporation condition}:
\begin{equation}\label{morsel-time-condition}
    \frac{\tau}{t_{\rm evap}} > 1 \implies \frac{\mu}{m} < \left(\frac{32 G(M-m) }{\gamma \, a^2 m } \right)^{1/2} \left( 1+\frac{4 G^2 (M-m)^2 }{a^2} \right) \left(1-\frac{8 G^2 (M-m)^2}{a^2} \right)^{-1/4} \ .
\end{equation}
\begin{figure}
    \centering
    \includegraphics[width=\linewidth]{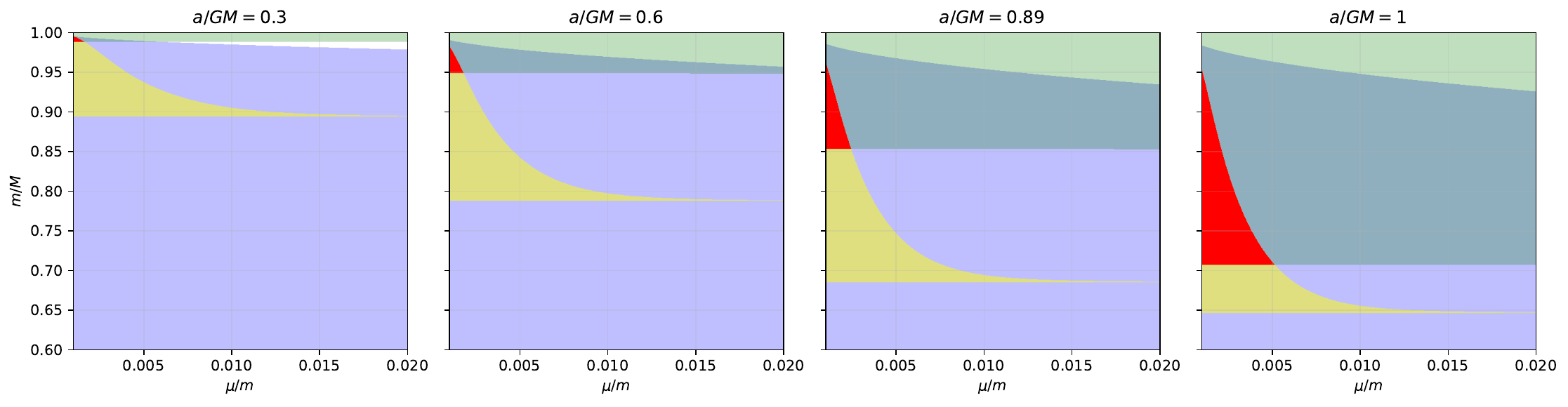}
    \caption{Regions in the parameter space $\mu/m$ vs. $m/M$ where the smaller fragment can evaporate, shown for initial mass $M = 10\ M_{\rm P}$ and $p=-1/2$. The shaded areas correspond to the domains where the following three inequalities are individually satisfied: yellow for the evaporation condition in Eq.~\eqref{morsel-time-condition}, blue for the velocity condition in Eq.~\eqref{condition on velocity} and green for the entropy condition in Eq.~\eqref{condition on entropy}. The red region highlights their simultaneous intersection.}
    \label{fig-flying-time}
\end{figure}
In Fig.~\ref{fig-flying-time} we illustrate the region (in red) compatible with the formation of a fragment, Eqs.~\eqref{condition on velocity} and~\eqref{condition on entropy}, and the condition for flying time larger than the evaporation time, Eq.~\eqref{morsel-time-condition}, for $p=-1/2$, $M=10\ M_{\rm P}$ and different values of the initial rotational parameter $a$. The plots show that it is possible for the fragment to evaporate before reabsorption, however only for masses close to the Planck one. For larger masses of the initial Kerr black hole, one can check that the yellow region corresponding to the evaporation condition in Eq.~\eqref{morsel-time-condition} shifts towards smaller $m/M$, hence the overlap ceases.

\subsection{Higher dimensions}\label{sec:higherdim}
The conditions for the stability of horizons change when GR is extended to more than four space-time dimensions. The reason is twofold: new extended geometries are allowed, such as strings and branes; black hole rotation velocities are not limited, as the gravitational attraction weakens with increasing dimensionality while the centrifugal barrier does not. In fact, in higher dimensional GR, both black strings and ultra-spinning black holes are known to be subject to fragmentation \cite{Gregory:1993vy,Emparan:2014jca}. We will briefly review these results, and sketch in a simple case how this type of fragmentation could be observable.

Let us start with a black string, whose metric is~\cite{Horowitz:1991cd}:
\begin{equation}
    \dd s^2=-f(r) \dd t^2+f(r)^{-1} \dd r^2+r^2 \dd \Omega_2^2 + \dd z_i^2
\end{equation}
with $f(r)=1-2GM/r$, $M$ being the mass of the string, and $z_i$ ($i=4,\dots d$) the coordinates of the extra dimensions.
It can be thought of as an ordinary Schwarzschild black hole with radius $2GM$ stretched along the extra spatial dimensions. Black strings of this type are subject to the Gregory--Laflamme~\cite{Gregory:1993vy} instability for large enough extra dimension, i.e.  with volume much larger than the scale given by the mass $M$ of the string.
When such an object is sufficiently long and thin, it becomes unstable to perturbations that vary along the extended direction. Small fluctuations with wavelengths longer than a critical value—typically of the order of the horizon radius—do not decay, instead they grow exponentially in time.
\begin{comment}The instability is closely related to thermodynamic considerations: thin black strings have a smaller entropy with respect to black holes with the same mass.
As the instability develops, the initially uniform black string is expected to evolve toward a non-uniform configuration, possibly developing into a sequence of black holes connected by thin necks.\end{comment} 
{The instability admits a thermodynamic heuristic: at fixed mass, a uniform string longer than a critical wavelength $\lambda_{\rm GL}\sim r_h$ has smaller horizon area than a localized (caged) black hole of the same mass, so the uniform string is entropically disfavored. Dynamically, the linearized horizon perturbations with wavelength $\lambda>\lambda_{\rm GL}$ do not redshift away but grow exponentially, $\delta r_h\sim e^{\Omega t}$ with $\Omega>0$ below the threshold wavenumber~\cite{Gregory:1993vy}. The horizon is thereby driven toward the inhomogeneous ``beads-on-a-string'' morphology of spherical bulges joined by thin necks, whose subsequent fate is a genuinely dynamical question that has been addressed numerically\cite{Gregory:1993vy, Lehner:2010pn}.}

Ultra-spinning black holes approach the geometry of black strings, and this is why they also suffer from instability ~\cite{Emparan:2003sy,Emparan:2008eg}.
For simplicity, let us consider a black hole in a $d$-dimensional spacetime, with total mass $M$ and total angular momentum $J$ rotating only in the plane $(\theta,\varphi)$, with coordinates $(t,r,\theta,\theta_1, \ldots ,\theta_{d-4},\varphi)$: its metric is described by the Myers--Perry metric \cite{Myers:1986un}, which reads
\begin{align}\label{myersperrymetric}
    \dd s^2 &= - \dd t^2 + \frac{\mu}{R^{d-5}\Sigma}\left(\dd t + a \sin^2 \theta \dd \varphi\right)^2  + \frac{\Sigma}{\Delta}\dd R^2 + \notag \\ & \qquad  + \Sigma \, \dd \theta^2 + (R^2+ a^2)\sin^2 \theta \dd \varphi^2 + R^2 \cos^2 \theta \, \dd \Omega_{d-4}^2\ ,
\end{align}
where $\dd \Omega_N^2$ is the line element on the sphere $\mathbb{S}^{N}$
\begin{equation}
    \dd \Omega_{N}^2 =  \sum_{i=0}^{N-1} \prod_{j=1}^i\sin^2 \theta_j \dd \theta_{i+1}^2 \ ,
\end{equation}
where $\{ \theta_i \}_{1\leq i\leq d-2}$ represent the angular coordinates on the unit sphere $\mathbb{S}^{d-2}$, with $\theta_i \in [0,\pi]$ for $i=1, \ldots, d-3$ and $\theta_{d-2} \in [0,2\pi[$. Moreover, we defined the functions
\begin{equation}
    \Sigma=R^2 +a^2 \cos^2 \theta \qq{and} \Delta=R^2+a^2- \mu \, R^{5-d} \ ,
\end{equation}
in which the parameters $\mu$ and $a$ are respectively related to the mass-radius $r_M$ and the rotational radius, defined as
\begin{equation}\label{higherdim-gravradius-rotradius}
    \mu=r_M^{d-3}=\frac{16 \pi G_d M}{(d-2)\Omega_{d-2}} \qq{and} a= \frac{d-2}{2}\frac{J}{M} \ 
\end{equation}
where $\Omega_{d-2}$ is the area of the sphere $\mathbb{S}^{d-2}$ and $G_d$ is the $d$-dimensional Newton's gravitational constant.
As discussed in~\cite{Emparan:2003sy}, a heuristic explanation of the competition between the gravitational force and the centrifugal repulsion can be obtained by analyzing the following expression:
\begin{equation}
    \frac{\Delta}{R^2} -1 = -\frac{\mu}{R^{d-3}} + \frac{a^2}{R^2} \ .
\end{equation}
Consequently, we observe that as the number of dimensions increases, the gravitational interaction represented by the first term diminishes, whereas the centrifugal barrier in the second term remains independent of $d$. This property significantly influences the characteristics of the horizon, which is defined as the largest root of $g_{rr}^{-1}$, specifically 
\begin{equation} \label{higherdim-horizon}
    R_+^2 + a^2 - \mu\,  R_+^{5-d} =0
\end{equation}

\begin{figure}
    \centering

% Gradient Info
  
\tikzset {_78saf8vmo/.code = {\pgfsetadditionalshadetransform{ \pgftransformshift{\pgfpoint{108.29 bp } { -137.41 bp }  }  \pgftransformscale{1.82 }  }}}
\pgfdeclareradialshading{_lfdcl0d63}{\pgfpoint{-72bp}{88bp}}{rgb(0bp)=(1,1,1);
rgb(0bp)=(1,1,1);
rgb(25bp)=(0,0,0);
rgb(400bp)=(0,0,0)}

% Gradient Info
  
\tikzset {_100k1qlw9/.code = {\pgfsetadditionalshadetransform{ \pgftransformshift{\pgfpoint{194.4 bp } { -164.4 bp }  }  \pgftransformscale{1.2 }  }}}
\pgfdeclareradialshading{_5yzqaxrsj}{\pgfpoint{-176bp}{152bp}}{rgb(0bp)=(1,1,1);
rgb(0bp)=(1,1,1);
rgb(25bp)=(0,0,0);
rgb(400bp)=(0,0,0)}

% Gradient Info
  
\tikzset {_be8s90dbf/.code = {\pgfsetadditionalshadetransform{ \pgftransformshift{\pgfpoint{108.29 bp } { -137.41 bp }  }  \pgftransformscale{1.82 }  }}}
\pgfdeclareradialshading{_fyfnemd5w}{\pgfpoint{-72bp}{88bp}}{rgb(0bp)=(1,1,1);
rgb(0bp)=(1,1,1);
rgb(25bp)=(0,0,0);
rgb(400bp)=(0,0,0)}
\tikzset{every picture/.style={line width=0.75pt}} %set default line width to 0.75pt        

\begin{tikzpicture}[x=0.75pt,y=0.75pt,yscale=-1,xscale=1]

%Shape: Circle [id:dp5234531563174398] 
\draw  [draw opacity=0][shading=_lfdcl0d63,_78saf8vmo] (279,78.4) .. controls (279,68.13) and (287.33,59.8) .. (297.6,59.8) .. controls (307.87,59.8) and (316.2,68.13) .. (316.2,78.4) .. controls (316.2,88.67) and (307.87,97) .. (297.6,97) .. controls (287.33,97) and (279,88.67) .. (279,78.4) -- cycle ;
%Shape: Ellipse [id:dp24215066346449932] 
\draw  [draw opacity=0][shading=_5yzqaxrsj,_100k1qlw9] (22.49,122.25) .. controls (18.02,103.75) and (40.62,82.4) .. (72.99,74.57) .. controls (105.36,66.73) and (135.23,75.37) .. (139.71,93.87) .. controls (144.18,112.36) and (121.58,133.71) .. (89.21,141.55) .. controls (56.84,149.38) and (26.97,140.74) .. (22.49,122.25) -- cycle ;
%Straight Lines [id:da4253954681235512] 
\draw [line width=1.5]    (60.03,34.86) -- (73.2,84.2) ;
\draw [shift={(59,31)}, rotate = 75.06] [fill={rgb, 255:red, 0; green, 0; blue, 0 }  ][line width=0.08]  [draw opacity=0] (13.4,-6.43) -- (0,0) -- (13.4,6.44) -- (8.9,0) -- cycle    ;
%Curve Lines [id:da9360789386465044] 
\draw    (73.68,93.78) .. controls (31.44,108.07) and (25,82.2) .. (56.2,66.6) ;
\draw [shift={(77,92.6)}, rotate = 159.61] [fill={rgb, 255:red, 0; green, 0; blue, 0 }  ][line width=0.08]  [draw opacity=0] (7.14,-3.43) -- (0,0) -- (7.14,3.43) -- (4.74,0) -- cycle    ;
%Straight Lines [id:da15476044207621598] 
\draw  [dash pattern={on 4.5pt off 4.5pt}]  (154,100.4) -- (212,100.4) ;
\draw [shift={(215,100.4)}, rotate = 180] [fill={rgb, 255:red, 0; green, 0; blue, 0 }  ][line width=0.08]  [draw opacity=0] (10.72,-5.15) -- (0,0) -- (10.72,5.15) -- (7.12,0) -- cycle    ;
%Shape: Circle [id:dp0850846658822687] 
\draw  [draw opacity=0][shading=_fyfnemd5w,_be8s90dbf] (248.6,124) .. controls (248.6,113.73) and (256.93,105.4) .. (267.2,105.4) .. controls (277.47,105.4) and (285.8,113.73) .. (285.8,124) .. controls (285.8,134.27) and (277.47,142.6) .. (267.2,142.6) .. controls (256.93,142.6) and (248.6,134.27) .. (248.6,124) -- cycle ;
%Straight Lines [id:da9587575886855132] 
\draw [line width=1.5]    (260.17,39.48) -- (297.6,78.4) ;
\draw [shift={(257.4,36.6)}, rotate = 46.12] [fill={rgb, 255:red, 0; green, 0; blue, 0 }  ][line width=0.08]  [draw opacity=0] (13.4,-6.43) -- (0,0) -- (13.4,6.44) -- (8.9,0) -- cycle    ;
%Straight Lines [id:da5217162845485434] 
\draw [line width=1.5]    (267.2,124) -- (304.63,162.92) ;
\draw [shift={(307.4,165.8)}, rotate = 226.12] [fill={rgb, 255:red, 0; green, 0; blue, 0 }  ][line width=0.08]  [draw opacity=0] (13.4,-6.43) -- (0,0) -- (13.4,6.44) -- (8.9,0) -- cycle    ;
%Straight Lines [id:da7710400141948585] 
\draw  [dash pattern={on 4.5pt off 4.5pt}]  (274.54,61.58) -- (241.96,92.92) ;
\draw [shift={(239.8,95)}, rotate = 316.11] [fill={rgb, 255:red, 0; green, 0; blue, 0 }  ][line width=0.08]  [draw opacity=0] (7.14,-3.43) -- (0,0) -- (7.14,3.43) -- (4.74,0) -- cycle    ;
\draw [shift={(276.7,59.5)}, rotate = 136.11] [fill={rgb, 255:red, 0; green, 0; blue, 0 }  ][line width=0.08]  [draw opacity=0] (7.14,-3.43) -- (0,0) -- (7.14,3.43) -- (4.74,0) -- cycle    ;
%Straight Lines [id:da4194885169064725] 
\draw  [dash pattern={on 0.84pt off 2.51pt}]  (331,189.4) -- (215.8,71.8) ;
%Straight Lines [id:da6705788822080009] 
\draw  [dash pattern={on 0.84pt off 2.51pt}]  (360.6,142.2) -- (245.4,24.6) ;

% Text Node
\draw (38,33.2) node [anchor=north west][inner sep=0.75pt]    {$J$};
% Text Node
\draw (123.2,52.2) node [anchor=north west][inner sep=0.75pt]    {$M$};
% Text Node
\draw (321.6,55.8) node [anchor=north west][inner sep=0.75pt]    {$m$};
% Text  Node
\draw (238.4,137.4) node [anchor=north west][inner sep=0.75pt]    {$m$};
% Text Node
\draw (239.2,57.8) node [anchor=north west][inner sep=0.75pt]    {$2b$};

\end{tikzpicture}
    \caption{Fragmentation of an ultra-spinning black hole of mass $M$ and angular momentum $J$, into two non-rotating black holes with impact parameter $2b$.}
    \label{ultra-spinning-fragment}
\end{figure}
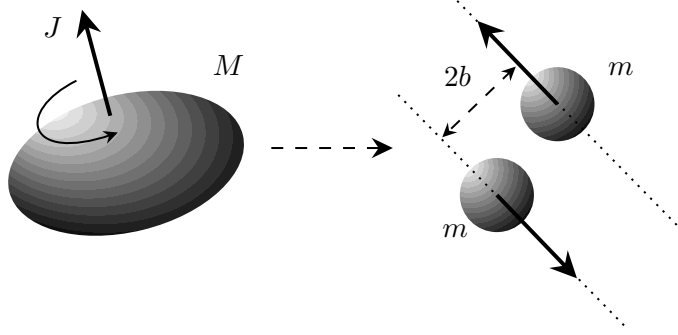

It has been shown that for ultra-spinning initial black holes, several fragmentation scenarios become entropically favored in dimensions $d>5$ \cite{Andrade:2020ilm}. Let us explicitly explain one of them~\cite{Emparan:2003sy} by considering an ultra-spinning black hole with mass $M$ and angular momentum $J$ fragmenting into two non-rotating black holes of equal mass $m$ flying apart in antiparallel directions with impact parameter $2b$. The total conserved energy and angular momentum are $M$ and $J$, so the momenta of the final black holes in the center-of-momentum frame are $\pm J / 2 b$, such that the on-shell conservation is
\begin{equation}
M=2 \sqrt{m^2+\frac{J^2}{4 b^2}}\ ,
\end{equation}
yielding the mass of an individual black hole fragment and their  horizon radii
\begin{equation}
    m=\frac{1}{2} \sqrt{M^2-\frac{J^2}{b^2}} \qq{and} r_m^{d-3}= \frac{16 \pi G m}{(d-2)\Omega_{d-2}}\ .
\end{equation}
Using the horizon condition in Eq.~\eqref{higherdim-horizon} and the definitions in Eq.~\eqref{higherdim-gravradius-rotradius}, we can relate the radius of the two fragments to the initial black hole's horizon radius as
\begin{equation}
    r_m^{d-3} = \frac{R_+^{d-5}(R_+^2+a^2)}{2}\sqrt{1 - \frac{4a^2}{(d-2)^2b^2}} \ .
\end{equation}
The area of the horizon of the initial Myers--Perry black hole, which is invariant under boosts~\cite{Horowitz:1997fr}, is computed as 
\begin{equation} \label{higherdim-horizon-area}
    \mathcal{A}_0=\int \sqrt{h}\, \dd^{d-2} \theta \ ,
\end{equation}
where $h$ is the determinant of the induced metric on the horizon that, from Eq.~\eqref{myersperrymetric}, reads
\begin{equation}
    h = \Sigma_+ \left( R_+^2+a^2  + \frac{\mu \, a^2 \sin^2 \theta}{R_+^{d-5}\Sigma_+} \right)\sin^2  \theta \, (R_+^2 \cos^2\theta)^{d-4} \det \Omega_{d-4} \ ,
\end{equation}
with $\Sigma_+ = R_+^2 + a^2 \cos^2 \theta$. Using the condition in Eq.~\eqref{higherdim-horizon}, we can simplify the bracket and integrate Eq.~\eqref{higherdim-horizon-area}, obtaining the horizon area for the rotating black hole
\begin{equation}\label{higherdim-initial-area}
    \mathcal{A}_0 = \Omega_{d-2}R_+^{d-4}(R_+^2+a^2) \ .
\end{equation}
The fragmentation becomes entropically favored if the total final area is larger than the initial one, so we compare Eq.~\eqref{higherdim-initial-area} with the sum of the areas of the two non-rotating BHs
\begin{equation}
    \frac{\mathcal{A}_1}{\mathcal{A}_0}=\frac{2 r_{m}^{d-2}}{ R_{+}^{d-4}\left(R_{+}^2+a^2\right)} = \left[\frac{1+(a/R_+)^2}{2}\left( 1-\frac{4 a^2}{(d-2)^2b^2}\right)^{\frac{d-2}{2}} \right]^{1/(d-3)}>1 \ .
\end{equation}
The reasonable range of values for the impact parameter $b$ is between the horizon radius and the rotational radius of the initial black hole $ a < b < \sqrt{R_+^2 + a^2}$ and in the ultra-spinning regime, the value of $\mathcal{A}_1/ \mathcal{A}_0$ is maximized, entropically favoring the fragmentation into two black holes.
%For every dimension $D>5$ there exists a critical value of the dimensionless spin parameter such that for $a/r_m>a_{\rm crit}>1$ (i.e.\ ultra-spinning regime, where the rotational radius exceeds the mass-radius) the final area (and thus the final entropy) is bigger than the initial one, signalling the possibility of fragmentation.
This instability, just as the Gregory--Laflamme instability, has been confirmed numerically \cite{Andrade:2019edf}.{\ We stress, however, that what the numerics establish is the onset and nonlinear development of the instability and the emergence of a beads-on-a-string morphology; the terminal necks reach a regime of diverging curvature (a classical naked singularity, in violation of cosmic censorship~\cite{Lehner:2010pn,Horowitz:2001cz}), and a rigorous demonstration that the endpoint is a regular set of black holes is not currently available.}

Having determined that horizons can fragment in higher dimensional space-time, one can wonder what an observer in four-dimensional spacetime would see during the fragmentation process. The answer depends significantly on the configuration of the extra dimensions, which we assume to be compact over volumes small enough to be unobservable. The simplest configuration is one in which the observers are localized on a 4D brane, as one of the boundaries of the extra dimensional volume.
Let us first consider a black string stretching out from the observer brane.
Since fragmentation occurs along the extra dimension, the localized observer would simply see the object's area decrease progressively. Consider now an ultra-spinning black hole in the $d$-dimensional space time, which may or may not intersect the observer's brane. Fragmentation occurs perpendicularly to the rotation axis, hence the appearance of fragments could  be directly visible for the observer.
Even when the fragmentation process takes place outside the brane, its gravitational effects can still be observable. To qualify this statement, in the next section we will discuss the influence of a $d$-dimensional black hole on a 4D brane.

\subsubsection{Induced metric of a $d$-dimensional black hole on a 4-dimensional brane}
Let us suppose that we (observers) live on a 4-dimensional brane in a universe with $d$ dimensions, which contains a static and spherical black hole, smaller than the sizes of the compact dimensions. This object does not necessarily intersect the 4-dimensional subspace in which we live. Hence, we start from a metric given by the Schwarzschild-Tangherlini form~\cite{Tangherlini:1963bw} in $d$-dimensional spherical coordinates:
\begin{equation}
    \dd s^2 = - f(R) \dd t^2 + f(R)^{-1} \dd R^2 + R^2 \dd \Omega_{d-2}^2 \ .
\end{equation}
The metric lapse function depends only on the $d$-dimensional radius $R$ and reads
\begin{equation}
 f(R)=1-\frac{\mu}{R^{d-3}} \ ,
\end{equation} 
where $\mu$ is related to the horizon radius $R_H = \mu^{1/(d-3)}$.

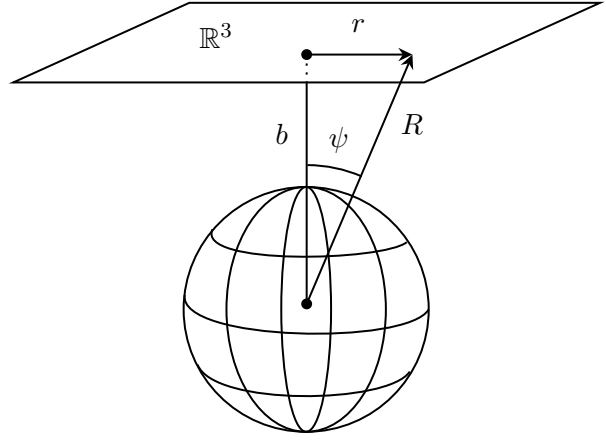
\begin{figure}
    \centering

\tikzset{every picture/.style={line width=0.75pt}} %set default line width to 0.75pt        

\begin{tikzpicture}[x=0.75pt,y=0.75pt,yscale=-1,xscale=1]

%Shape: Circle [id:dp639923301123618] 
\draw   (93.5,225) .. controls (93.5,191.03) and (121.03,163.5) .. (155,163.5) .. controls (188.97,163.5) and (216.5,191.03) .. (216.5,225) .. controls (216.5,258.97) and (188.97,286.5) .. (155,286.5) .. controls (121.03,286.5) and (93.5,258.97) .. (93.5,225) -- cycle ;
%Shape: Ellipse [id:dp034858946590230144] 
\draw   (115,225) .. controls (115,191.03) and (132.91,163.5) .. (155,163.5) .. controls (177.09,163.5) and (195,191.03) .. (195,225) .. controls (195,258.97) and (177.09,286.5) .. (155,286.5) .. controls (132.91,286.5) and (115,258.97) .. (115,225) -- cycle ;
%Shape: Ellipse [id:dp1902717231129436] 
\draw   (142.5,225) .. controls (142.5,191.03) and (148.1,163.5) .. (155,163.5) .. controls (161.9,163.5) and (167.5,191.03) .. (167.5,225) .. controls (167.5,258.97) and (161.9,286.5) .. (155,286.5) .. controls (148.1,286.5) and (142.5,258.97) .. (142.5,225) -- cycle ;
%Curve Lines [id:da10722862043991799] 
\draw    (94,217.75) .. controls (91,241.25) and (206,243.25) .. (216.5,225) ;
%Curve Lines [id:da6625518818971505] 
\draw    (108,184.75) .. controls (99.5,203.25) and (193.5,200.75) .. (205.5,191.25) ;
%Curve Lines [id:da8549642835911464] 
\draw    (100.5,252.75) .. controls (108.5,273.75) and (196,272.75) .. (207,256.25) ;
%Shape: Parallelogram [id:dp8714327592380082] 
\draw   (96.35,71) -- (302.5,71) -- (214.15,111) -- (8,111) -- cycle ;
%Straight Lines [id:da1664019892315609] 
\draw    (155.25,111) -- (155.25,222.25) ;
%Shape: Circle [id:dp5878575392434392] 
\draw  [fill={rgb, 255:red, 0; green, 0; blue, 0 }  ,fill opacity=1 ] (153.02,222.25) .. controls (153.02,221.02) and (154.02,220.02) .. (155.25,220.02) .. controls (156.48,220.02) and (157.48,221.02) .. (157.48,222.25) .. controls (157.48,223.48) and (156.48,224.48) .. (155.25,224.48) .. controls (154.02,224.48) and (153.02,223.48) .. (153.02,222.25) -- cycle ;
%Shape: Circle [id:dp42974202631699165] 
\draw  [fill={rgb, 255:red, 0; green, 0; blue, 0 }  ,fill opacity=1 ] (153.02,97) .. controls (153.02,95.77) and (154.02,94.77) .. (155.25,94.77) .. controls (156.48,94.77) and (157.48,95.77) .. (157.48,97) .. controls (157.48,98.23) and (156.48,99.23) .. (155.25,99.23) .. controls (154.02,99.23) and (153.02,98.23) .. (153.02,97) -- cycle ;
%Straight Lines [id:da2276041903324878] 
\draw  [dash pattern={on 0.84pt off 2.51pt}]  (155.25,97) -- (155.25,111) ;
%Straight Lines [id:da3741315768849899] 
\draw    (155.25,97) -- (205.33,97) ;
\draw [shift={(208.33,97)}, rotate = 180] [fill={rgb, 255:red, 0; green, 0; blue, 0 }  ][line width=0.08]  [draw opacity=0] (7.14,-3.43) -- (0,0) -- (7.14,3.43) -- (4.74,0) -- cycle    ;
%Straight Lines [id:da7634129110706215] 
\draw    (155.25,222.25) -- (207.16,99.76) ;
\draw [shift={(208.33,97)}, rotate = 112.97] [fill={rgb, 255:red, 0; green, 0; blue, 0 }  ][line width=0.08]  [draw opacity=0] (7.14,-3.43) -- (0,0) -- (7.14,3.43) -- (4.74,0) -- cycle    ;
%Shape: Arc [id:dp36824261729931995] 
\draw  [draw opacity=0] (155.66,152.52) .. controls (165.3,152.58) and (174.48,154.59) .. (182.82,158.18) -- (155.25,222.25) -- cycle ; \draw   (155.66,152.52) .. controls (165.3,152.58) and (174.48,154.59) .. (182.82,158.18) ;  

% Text Node
\draw (138.11,130.29) node [anchor=north west][inner sep=0.75pt]    {$b$};
% Text Node
\draw (176.33,78.07) node [anchor=north west][inner sep=0.75pt]    {$r$};
% Text Node
\draw (201,125.29) node [anchor=north west][inner sep=0.75pt]    {$R$};
% Text Node
\draw (164.89,131.84) node [anchor=north west][inner sep=0.75pt]    {$\psi $};

\draw (100,80) node [anchor=north west][inner sep=0.75pt]    {$\mathbb{R}^3$};

\end{tikzpicture}

    \caption{Embedding of $\mathbb{R}^3$ in $\mathbb{R}^{d-1}$ where a higher-dimensional BH is present. Here, $b$ is the distance of the manifold from the center of the BH in the orthogonal space and $r$ the projected distance along $\mathbb{R}^3$.}
    \label{fig:brane-embedding}
\end{figure}

We can consider the equal time space slice, which returns the Riemannian metric in spherical coordinates
\begin{equation}
    \dd s^2|_{t \, {\rm fixed}} =  f(R)^{-1} \dd R^2 + R^2 \dd \Omega_{d-2}^2 \ ,
\end{equation}
in which we recognize our 3-dimensional subset, namely $\mathbb{R}^{d-1} \supset \mathbb{R}^{3} \times \mathbb{R}^{d-4}$. Measuring the distance between the black hole center and the observers' brane as an impact parameter $b$, we can write
\begin{equation} \label{Rdsquared}
    R^2 = r^2 + b^2 \ .
\end{equation}
We can take now $\psi \equiv \theta_1$ as the latitude angle from the axis perpendicular to the $\mathbb{R}^3$ hypersurface, so $r=R \sin \psi$ and $b = R \cos \psi$. From the latter relation, we have
\begin{equation}
    0=\dd (R \cos \psi) = \cos \psi \, \dd R - R \sin \psi \, \dd \psi \implies \dd \psi = \frac{1 }{R \tan \psi} \dd R \ ,
\end{equation}
and from Eq.~\eqref{Rdsquared}, we have
\begin{equation}
    \dd R = \frac{r}{\sqrt{r^2+b^2}}\dd r \ ,
\end{equation}
which, together with $\tan \psi = r/b$, yields
\begin{equation}
    \dd \psi = \frac{b}{r^2+ b^2} \dd r \ .
\end{equation}
The angular part is now factorized as
\begin{equation}
    S^{d-2} \mapsto S^2 \times S^{d-5} \implies \dd \Omega^2_{d-2} = \dd\psi^2+\sin^2 \psi \dd \Omega_{2}^2 + \cos^2 \psi \dd \Omega_{d-5}^2 \ .
\end{equation}
The resulting metric is given in the warped form
\begin{equation}\label{5dtangherlinireduced}
    \dd s^2 = -f(r)\dd t^2 + \left(\frac{r^2}{f(r)}+b^2 \right)\frac{\dd r^2}{r^2 + b^2} + r^2 \dd \Omega_2^2 + b^2\dd \Omega_{d-5}^2 \ ,
\end{equation}
which tells about a factorized spacetime as $\mathcal{M}_4 \times S_b^{d-5}$ (one dimension is ``missing'' from the embedding), where the lapse function reads
\begin{equation}
    f(r) = 1- \frac{\mu}{(r^2+b^2)^{(d-3)/2}} \ .
\end{equation}
It is clear that the 4-dimensional metric is not a solution to the vacuum Einstein equations in 4 dimensions, as it differs from the Schwarzschild solution. The presence of the black hole with mass $\propto \mu^{1/(d-3)}$ in the bulk is effectively affecting the geometry of the subspace, even if the brane does not intersect the horizon, depending on its distance $b$, while the ``gravitational potential" depends on the number of dimensions $d$. Taking the distance $b=0$, corresponding to the equatorial form of the bulk metric, Eq.~\eqref{5dtangherlinireduced} reduces to the 4$d$ Schwarzschild form, while in the limit $b\to \infty$ the Ricci scalar scales as $\sim b^{-8}$ recovering the flat spacetime.

We can generalize this argument to a rotating black hole, whose metric is given in Eq.~\eqref{myersperrymetric}. Let us consider a $d=5$ Myers--Perry black hole with a single rotation around the angle $\varphi$, with metric
\begin{equation}
    \dd s^2 = - \dd t^2 + \frac{\mu}{\Sigma}\left( \dd t + a \sin^2\theta \dd \varphi\right)^2 + \frac{\Sigma}{\Delta} \dd R^2 + \Sigma \dd \theta^2 + ( R^2 + a^2)\sin^2 \theta \dd\varphi^2 + R^2 \cos^2 \theta \dd \psi^2 \ .
\end{equation}
Placing the rotation plane into the 4$d$ brane means that in the extra dimension there is no rotation and that the induced metric inherits a warped form as
\begin{multline}
    \dd s^2 = - \left(1 - \frac{\mu}{\Sigma} \right) \dd t^2 + \frac{2 \mu a \sin^2 \theta}{\Sigma} \dd t \dd \varphi + \left( r^2 + b^2 + a^2 + \frac{\mu a^2 \sin^4 \theta}{\Sigma} \right)\dd \varphi^2 + \\ +\frac{r^2}{r^2+b^2}\frac{\Sigma}{\Delta} \dd r^2 + \Sigma \dd \theta^2+ (r^2 + b^2) \cos^2 \theta \dd \psi^2 \ ,
\end{multline}
where we have
\begin{equation}
    \Sigma = r^2 + b^2+a^2 \cos^2 \theta \qq{and} \Delta = r^2 + b^2 + a^2 - \frac{\mu}{r^2 + b^2 } \ .
\end{equation}
The 4-dimensional part of the metric resembles the Kerr form, with off diagonal elements that reproduce the frame dragging and effective radius $r^2 + b^2 + a^2 \cos^2\theta$.

In the case in which the rotation takes place fully in the extra dimensions, then the metric assumes the following warped form
\begin{equation}
    \dd s^2 = -f_{\rm eff}(r) \dd t^2 + \left(\frac{r^2}{f_{\rm eff}(r)}+b^2\right) \frac{\dd r^2}{r^2+b^2} + r^2 \dd \Omega_2^2 + b^2 \dd \psi^2\ ,
\end{equation}
where now we have
\begin{equation}
    f_{\rm eff}(r) = 1 - \frac{\mu}{r^2 + b^2 + a^2 \cos^2 \psi_0} \ ,
\end{equation}
in which the fixed extra dimensional angle $\psi_0$ encodes the orientation of our brane with respect to the rotation plane of the bulk and causes an effective shift in the lapse function but the rotation of the black hole is not visible.

\section{Effective fragmentation from superradiance instability?}
\label{sec:superradiance}  
While analyzing the (im)possibility of fragmentation for a Kerr black hole in four dimensions, we saw how the area law allows for the extraction of rotational energy from it. For the same reason, the area law also implies the possibility of superradiant scattering~\cite{Press:1972zz}, namely the amplification of waves interacting with a rotating black hole. For this effect to occur, bosonic particles with low masses are required, such that their Compton wavelength is of the order of the black hole radius. More precisely, for a mode of frequency $\omega$ and azimuthal number $m$, amplification occurs when the superradiant condition $\omega < m \Omega_H$ is satisfied, where $\Omega_H$ is the angular velocity of the horizon. In this regime, the reflected wave carries more energy than the incident one, with the excess energy extracted from the black hole rotational degrees of freedom. This process relies on the existence of the ergosphere, a region outside the horizon where negative-energy states (as measured at infinity) are allowed. It can be regarded as the wave analogue~\cite{Zeldovich:1971ffh} of the Penrose process~\cite{Penrose:1971uk}: an infalling particle can split into two fragments, one of which falls into the black hole with negative energy, while the other escapes to infinity with energy larger than the initial one. As a result, energy and angular momentum are effectively extracted from the black hole. 

It is well known that this mechanism can trigger instabilities in the presence of a trapping mechanism~\cite{Brito:2014wla,Brito:2015oca}. If the amplified radiation is confined and repeatedly scattered back onto the black hole, it undergoes successive amplifications, leading to an exponential growth of the field amplitude—the so-called black-hole bomb. Such trapping can arise in different physical contexts, for instance due to effective mass terms for the fields or environmental effects such as surrounding plasma, which generate a confining potential.   In standard setups, the trapping is typically axisymmetric, and the endpoint of the instability remains axisymmetric as well, usually corresponding to a black hole with modified mass and spin.

However, if the trapping mechanism has a more complex, non-axisymmetric geometry, the amplification process may become spatially non-uniform. In such a scenario, one can envisage what we may call, in analogy with optical systems, fragmented superradiance \cite{PhysRevLett.118.013603}. The extracted energy could accumulate in localized regions, potentially forming long-lived, spatially separated structures. If sufficiently efficient, this process may effectively generate black hole companions, leading to an apparent or effective fragmentation of the original object, even though the horizon remains topologically connected. 
This speculative non-axisymmetric, non-uniform, trapping may occur in specific physical scenarios, for instance caused by an effective mass induced by the interaction of photons or of some new scalar degrees of freedom with the plasma of the accretion disk \cite{Conlon:2017hhi, Dima:2020rzg, Lingetti:2022psy,Lambiase:2025twn}. In these cases, the geometry of the trapping depends on the distribution of the plasma, which could reasonably present anisotropies. 

The main obstacle to this effective fragmentation is represented by the low frequencies required by efficient superradiance:  these modes cannot be trapped in small regions, as they are limited by the extension of their wavelength.

Nevertheless, assuming a trapping in a small spherical region of radius $r_{trap}$ at a distance $d_{trap}$ from the horizon, we can estimate the maximum mass that can be trapped as:

\begin{equation}
    M_{trap}=\frac{\pi r_{trap}^2}{4 \pi d_{trap}^2} \epsilon E_{rot}\,,
\end{equation}
where $\epsilon$ is the superradiance efficiency at the wavelength $\lambda \sim r_{trap}$. A black hole is formed if $G M_{trap}> r_{trap}/2$.

Of course this is a very speculative reasoning and a full evolution of superradiant fields in this highly non-axisymmetric trapping scenario should be performed to really assess the possibility and efficiency of such process. For example, in the case of photon-plasma induced mass the confinement efficiency is strongly affected by non-linear effects that cause plasma blowout \cite{Cardoso:2020nst,Cannizzaro:2023ltu}.

\section{Black holes do wear pants beyond GR}
\label{sec:modified}

{The \textit{classical} area theorem relies on the null energy condition (NEC). Once semiclassical or quantum effects are included this assumption can fail: in Hawking evaporation a NEC-violating (negative-energy) flux crosses the horizon and the horizon area decreases, so the \textit{classical} area law is evaded~\cite{Hawking:1975vcx}. This is fully consistent with the generalized second law for $S_{\rm gen}=A/4G+S_{\rm matter}$.} We will now explore how classical deviations from GR can also modify the area law, entropically favoring black hole fragmentation. The first case consists in a class of higher derivative corrections to GR. In these theories, fragmentation is allowed for small enough black holes as compared to an intrinsic new scale encoded in the corrections. 
As a second attempt, we will investigate a more general case by analyzing what type of modifications to the black hole area and/or to entropy can entropically favor fragmentation. 
As it was the case for higher dimensional scenarios, an entropic study is not sufficient to really prove the instability as it can only provide a strong hint that this process can happen. This suggestion should be confirmed by dynamical studies that investigate how the black holes behave under perturbations. These studies have been performed only in quadratic gravity \cite{Held:2022abx,Lu:2017kzi,Brito:2013wya} and for higher-dimensional scenarios \cite{Gregory:1993vy, Lehner:2010pn}, confirming the fragmentation instability in these cases.
Finally, we will consider the recent example of `bumpy horizon' solutions, triggered by vorticity.

\subsection{Modified gravity: higher curvature corrections}

In alternative theories of gravity, the area law can be modified. This can happen either because the horizon radius, and thus the area, of a black hole is different from GR or because the entropy in such a theory does not equal the area of the horizon.
As a specific example, we will focus on theories of gravity modified by higher order terms, and show in which cases fragmentation may be entropically favored.

There exists a particular family of higher curvature gravity that admits simple extensions of the Schwarzschild solution \cite{Bueno:2016lrh,Bueno:2017qce,Bueno:2024dgm}. The Lagrangian of this family of modified GR can be written as:
\begin{equation}
    \mathcal{L}=\frac{1}{16 \pi G}\left[R+\sum_{n=3}^{\infty} \frac{\lambda_n}{M_c^{2(n-1)}} 
    \mathcal{R}_{(n)}\right]
    \label{lagrangian}
\end{equation}
where $M_c$ is some new energy scale, $\lambda_n$ are dimensionless couplings and $n$ is the order in curvature of each invariant $\mathcal{R}_{(n)}$, which will be formed by contractions of the metric and the Riemann tensor (but not its covariant derivatives).
The strongest constraint on the new physics scale $M_c$ currently comes from the ringdown observation of stellar mass black holes. In fact, higher-curvature operators shift the Quasi-Normal Mode (QNM)~\cite{cano:2024ezp, cano:2020cao, hussain:2022ins, wagle:2023fwl} spectrum away from the Kerr one. Since astrophysical black holes have $M\sim 10\text{--}100~\Msun$, the probed scale $\ell$ is of the order of their horizon scale 
$ r_h\sim 10\text{--}100~\text{km}$. The strongest published bound from this channel has been derived in the context of cubic and quartic gravity~\cite{Silva:2022srr}:
\begin{equation}
\;\ell\;\lesssim\; \mathcal O(30\text{--}50)\,\mathrm{km}
\;\Longleftrightarrow\;
M_c\;\gtrsim\;\mathcal O(4\text{--}7)\times 10^{-12}\,\text{eV}.\;
\label{eq:ringdownbound}
\end{equation}
In these theories, spherically symmetric vacuum solutions exist in the form
\begin{equation}
\dd s_f^2=-f(r) \dd t^2+f(r)^{-1} \dd r^2+r^2 \dd \Omega_{2}^2 \ ,
\end{equation}
with the function $f$ determined by the following differential equation:
\begin{equation}
\begin{gathered}
2 G M-(1-f) r=-\sum_{n=3}^{\infty} \frac{\lambda_n}{M_c^{2(n-1)}}\left(\frac{f^{\prime}}{r}\right)^{n-3}\left[\frac{f^{\prime 3}}{n}+\frac{(n-3) f+2}{(n-1) r} f^{\prime 2} \right. \\ \left. -\frac{2}{r^2} f(f-1) f^{\prime}-\frac{1}{r} f f^{\prime \prime}\left(f^{\prime} r-2(f-1)\right)\right]\,.
\end{gathered}
\end{equation}
Imposing asymptotic flatness and assuming the existence of a regular horizon, the solution is characterized by the horizon radius $r_h$ (defined by $f(r_h)=0$) and the Hawking temperature $T\equiv f'(r_h)/(4\pi)$ via the following implicit equations:
\begin{align}
2 G M=r_h-\sum_{n=3}^{\infty} \frac{\lambda_n(4 \pi T)^{n-1}}{M_c^{2 n-2} r_h^{n-2}} \frac{\left(2 n+(n-1) 4 \pi T r_h\right)}{n(n-1)}\,, \\
1=4 \pi T r_h+\sum_{n=3}^{\infty} \frac{\lambda_n(4 \pi T)^{n-1}}{M_c^{2 n-2} r_h^{n-1}} \frac{\left(2 n+(n-3) 4 \pi T r_h\right)}{n(n-1)}\,.
\end{align}
%% [Clarified: T is the Hawking temperature, r_h the horizon radius.]

Wald's entropy for these solutions is:
\begin{equation} \label{eq:WaldEntropy}
    \begin{aligned} & S=\frac{\pi r_h^2}{G}\left[1-2 \sum_{n=3}^{\infty} \frac{\lambda_n(4 \pi T)^{n-1}}{M_c^{2 n-2} r_h^{n-1}} \left(\frac{2}{(n-2) 4 \pi T r_h}+\frac{1}{n-1}\right)\right]+\frac{4 \pi}{G M_c^2} \sum_{n=3}^{\infty} \frac{\lambda_n \chi^{n-2}}{(n-2)}\,.\end{aligned}
\end{equation}
where $\chi$ is defined as the solution of the following equation:
\begin{equation}
    \sum_{n=3}^{\infty} \frac{2 \lambda_n \chi^{n-1}}{(n-1)} \equiv 1\ .
\end{equation}
As explained in \cite{Bueno:2017qce}, these relations are exact, as they are necessary conditions for having a smooth near-horizon geometry. The infinite sum present in these formulas is always convergent since for these solutions $\frac{T}{r_h}<1$ for every value of the black hole mass M and both the temperature and horizon radius go to 0 for $M \rightarrow0$. We have numerically checked that the convergence of the series holds, even for very small black hole masses.
\begin{figure}
    \centering
\includegraphics[width=0.7\linewidth]{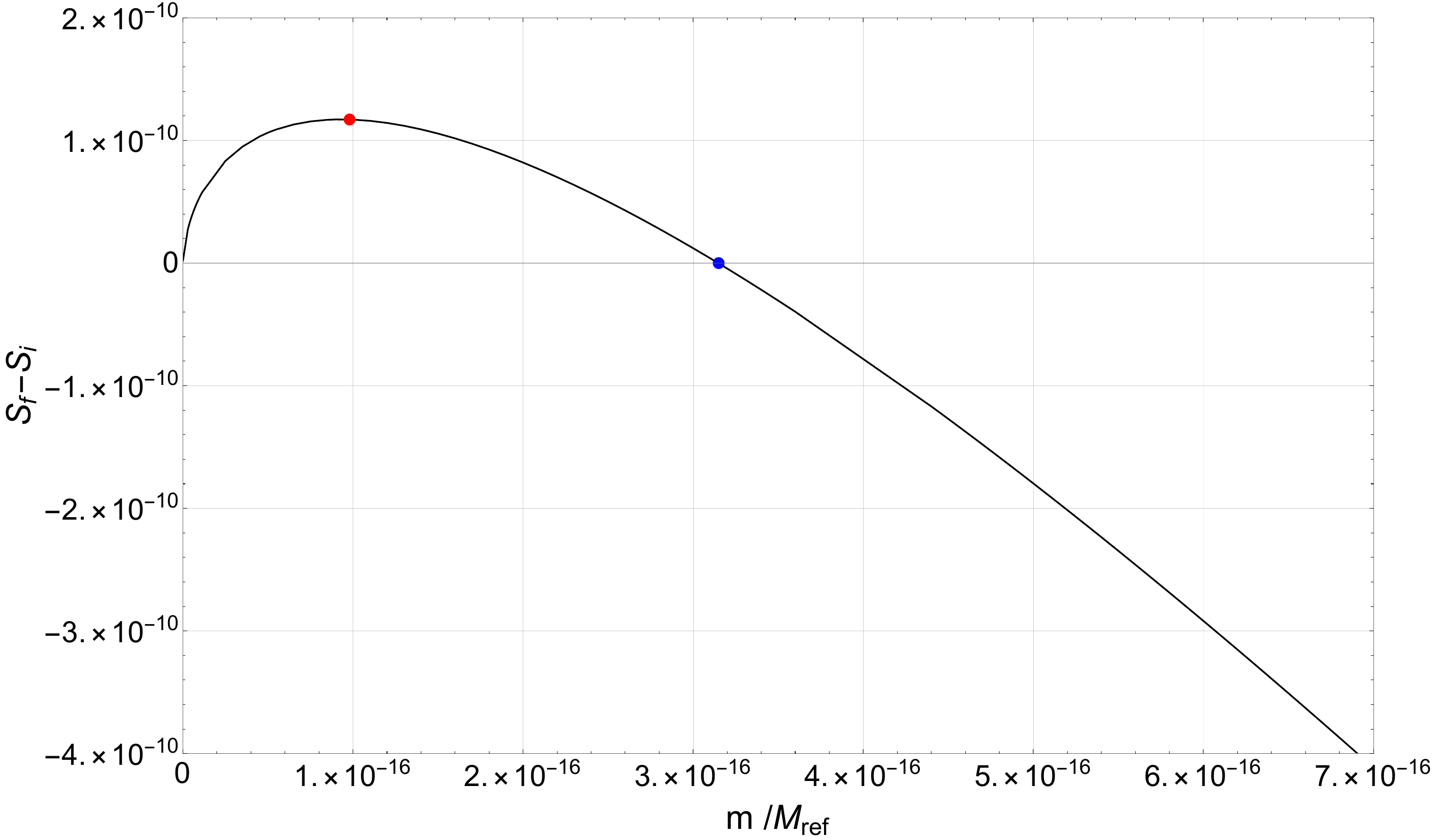}
    \caption{Change of entropy in the fragmentation of a black hole of initial mass $M_{ini}=10^{5} M_\text{ref}$ as a function of the mass of the smaller final fragment $m$. Here $M_\text{ref}$ is a reference mass scale built in the model, defined in Eq.~\eqref{eq:Mref}. The red dot represents the point where the fragmentation process leads to the largest increase in entropy, while the blue dot denotes the maximum mass that the smaller fragment can have while still keeping the fragmentation process entropically allowed ($S_f-S_i\geq0$). }
    \label{fig:dEntropy}
\end{figure}

The entropy in Eq.~\eqref{eq:WaldEntropy} can increase in the fragmentation process as long as one fragment is small enough, as the entropy is strongly enhanced for small black hole masses. It is interesting to notice that black hole mass dependence in the formulas above can always be reformulated in terms of the combination $M M_c/M_\text{P}^2 \equiv M/M_\text{ref}$, defining a reference mass scale as
\begin{equation} \label{eq:Mref}
    M_\text{ref} = \frac{M_\text{P}^2}{M_c} = 13\ M_\odot \times \left( \frac{10^{-11}~\text{eV}}{M_c}\right)\,.
\end{equation}
The numerical value $M_\text{ref} \sim 13\ M_\odot$ is the largest allowed by the direct bounds on $M_c$. In Fig.~\ref{fig:dEntropy} we show the difference in entropy between the initial and final state for a fragmentation process $M \to m + m_H$, where we assume $m \ll M$ (and $m_H = M-m$ from mass conservation) for a sample value $M = 10^5\ M_\text{ref}$. As anticipated, the fragmentation is entropically favored, i.e. $S_i - S_f < 0$, if one fragment is small enough, i.e. $m \lesssim 3.2 \times 10^{-16}\ M_\text{ref}$, while the maximal entropy points towards $m = 10^{-16}\ M_\text{ref}$. This result shows that for small enough $M_c$, fragmentation of astrophysical black holes may be possible: for instance, in our benchmark with $M = 10\ M_\odot$ (i.e., $M_\text{ref} = 10^{-4}\ M_\odot$), the most likely fragment mass would be $m \sim 10^{-20}\ M_\odot = 2\times 10^{10}~\text{kg}$.

\begin{figure}
    \centering
    \includegraphics[width=0.9\linewidth]{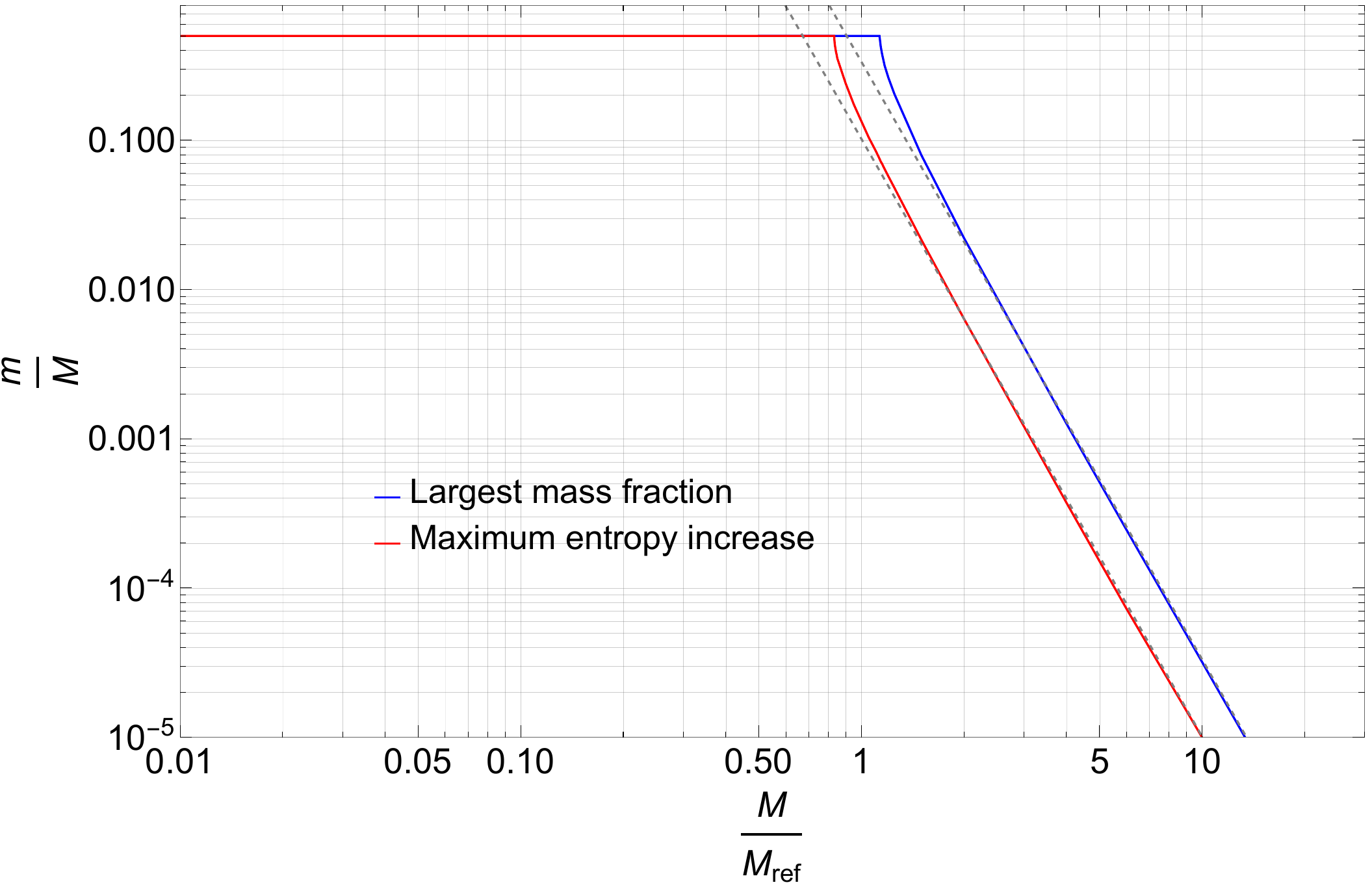}
    \caption{The largest allowed mass fraction of the smaller fragment is shown in blue, while the red line indicates the value that leads to the maximum increase of entropy during the fragmentation process. Both are shown as a function of the initial black hole mass $M$. For $M \gtrsim M_\text{ref}$, both curves are well fitted  by a $m/M \propto 1/M^4$ trend. }
    \label{fig:fragMass}

\end{figure}

In Fig.~\ref{fig:fragMass} we show the allowed range of fragment masses (below the blue line) and the most favored one (in red) as a function of the initial mass $M$. We see that as long as $M \gtrsim M_\text{ref}$, both the maximal and favored fragment mass scale like $M^{-3}$, more precisely:
\begin{equation}
    \frac{m_\text{max}}{M_\text{ref}} \sim  0.33\ \left( \frac{M}{M_\text{ref}} \right)^{-3}\,, \quad \frac{m_\text{fav.}}{M_\text{ref}} \sim  0.1\ \left( \frac{M}{M_\text{ref}} \right)^{-3}\,.
\end{equation}
For $M \lesssim M_\text{ref}$, instead, the entropy seems to favor fragmentation into equal mass fragments.

 \subsection{Modified gravity: General deformations}

Modifications to the horizon radius and/or to the area law generically emerge in many models affecting the geometry of black holes: for instance, solutions for regular black holes that have no singularity in their interior~\cite{Simpson:2018tsi,Hayward:2005gi,Ayon-Beato:1998hmi,bardeen_non-singular_1968,hayward_formation_2006,dymnikova_vacuum_1992,Fan:2016hvf}; solutions from theories of modified gravity, such as the one studied in the previous section; quantum corrected black holes~\cite{Binetti:2022xdi,DelPiano:2023fiw,DelPiano:2024gvw,DelPiano:2025ykr}; etc. 
If we commit to spherical symmetry, then the modification to the area law can be parametrized as
\begin{equation}
    S(M)= 4 \pi G M^2+h(M,\ell)\;\; \qq{(or equivalently, $S=4\pi\, M^2/M_{\rm P}^2 + h$),}
\end{equation}
where $h(M,\ell)$ is a generic function of the black hole mass $M$ and of a new distance scale parameter $\ell$, at which the new physics effects start to become important. In the quantum gravity framework the natural choice for $\ell$ is the Planck length $\ell \sim 1/M_\text{P}$. Usually the correction $h$ will be more important for small black hole masses (high curvature), as we would expect for corrections stemming from high energy modifications of GR.
An enhancement of the entropy for small black holes can, therefore, render the fragmentation thermodynamically allowed.

Let us analyze an explicit example to clarify this concept. Let us assume that the horizon radius is corrected as
\begin{equation}
r_h = 2 G M\  \left[1 + c_1 \left(\frac{\ell}{GM}\right)^{\alpha}+ \ldots \right] \ ,
\label{radius}
  \end{equation}
where $c_1$ is a dimensionless coefficient and $\ell$ denotes a length scale, and that the entropy is still given by one-quarter of the horizon area \begin{equation}
    S(M) = \pi \frac{r_h^2}{G} = 4 \pi G M^2 \  \left[1 + c_1 \left(\frac{\ell}{GM}\right)^{\alpha}+ \ldots \right]^2 \ .
\end{equation}
Then, the condition to have an increase of entropy in a fragmentation of a black hole of mass M into two fragments of masses $m$ and $M-m$
\begin{equation}
    S(M) < S(m) + S(M-m)
\end{equation}
can be satisfied if $m$ is small enough and $c_1 > 0$. However, typically the increase in entropy would require $G m \ll \ell$ for the smaller fragment, hence going deep in a regime where Eq.~\eqref{radius} may not hold.

It is interesting to note that, to our knowledge, all regular black hole models present in the literature seem stable under fragmentation, as for these models $c_1$ is generally a small negative constant~\cite{Simpson:2018tsi,Hayward:2005gi,Ayon-Beato:1998hmi,bardeen_non-singular_1968,hayward_formation_2006,dymnikova_vacuum_1992,Fan:2016hvf}. Intuitively, this is the case as, in order to regularize the central singularity, gravity should be weaker than in GR at the center of these objects. This implies that matter needs to go nearer the center $r=0$ in order to be trapped (i.e., to be inside the horizon) as compared to GR. Entropy constraints for these regular black holes allow for fragmentation only if $m<\ell$. However these models describe black holes (objects with at least a horizon) only if the mass is bigger than $k  \ell$ with a positive $k $.
We can conclude that all these models are stable under fragmentation if the entropy is still proportional to their area.\footnote{The same conclusions can be obtained without assuming the area law, through the first law of black hole thermodynamics if one assumes the ADM mass as estimator of the mass and the surface gravity as estimator of the temperature.}

A particularly interesting type of corrections to the black hole entropy is in the logarithmic form, since it was obtained in different scenarios such as loop quantum gravity and conformal field theory \cite{Carlip:2000nv} and in the effective field theory description \cite{DelPiano:2023fiw,DelPiano:2024gvw}. In these scenarios, the entropy is related to the black hole mass as
\begin{equation}
    S(M)=4 \pi G M^2 +\alpha \log(\ell^2/{(GM)}^2) + \ldots
    \label{entropy}
\end{equation}
For this type of correction, the function $h(M,\ell)$ blows up for $M\to 0$, thus to maintain a physically reasonable picture one must assume one of the two following features: either the mass of the black hole cannot decrease beyond a minimum value, for instance beyond the Planck mass, or other corrections to the entropy become more important for very small masses changing the behavior of $S$ near $M \sim 0$. In both cases the regime of validity of Eq.~\eqref{entropy} is bounded to masses $M$ bigger than a certain threshold value, which we can safely assume to be $M>\ell/G$. Furthermore, to ensure that the entropy is non-negative everywhere, the coefficient  $\alpha$ must be bounded by {$0\leq \alpha \leq 4\pi\, e\,\ell^2/G$: the upper limit stems from computing the entropy at the minimum $M_\ast=\sqrt{\alpha/4\pi G}$, where $S(M_\ast)=\alpha[\,1+\ln(4\pi\,\ell^2/(G\,\alpha))\,] \geq 0$}.
Subject to these constraints, one can verify that the quantity $\Delta=S(M)-S(M-m)-S(m)$ becomes negative for suitable values of $\alpha$, the fragment mass $m$ and $M$, with both $M$ and $m$ greater than $\ell/G$. 
Indeed, for $\alpha$ sufficiently large, but still within the allowed range, the logarithmic contribution dominates and $\Delta < 0$, meaning that entropy increases upon fragmentation. This occurs when the initial mass $M$ is small enough relative to $\ell$ that the concavity of the logarithmic correction prevails over the convexity of the leading $4GM^2$ term.

In general, with all this kind of entropy corrections, the possibility of fragmentation process is relevant only for small initial black hole masses. Indeed, for stellar mass black holes the fragmentation is possible only for fragments with masses smaller than the Planck mass for which we expect these effective corrections to the entropy to be no longer valid.
This means that we do not expect fragmentation to be relevant for an astrophysical scenario. However it can be relevant for the black hole evaporation process and for primordial black holes (PBHs), influencing their expected mass distribution and the constraints on the dark matter fraction in the form of PBHs.

{It is instructive to compare these entropic channels with the process that is always available in four dimensions: the semiclassical Hawking emission of a Planck-mass quantum \cite{Volovik:2025kao}, which by the hoop conjecture may itself be regarded as a (Planck-scale) black hole if it is sufficiently localized. For a parent of mass $M$ the Hawking temperature is $T_H=1/(8\pi G M)=M_{\rm P}^2/(8\pi M)$, so the emission of a quantum of energy $\omega\sim M_{\rm P}$ carries a Boltzmann suppression $\Gamma\sim\exp(-\omega/T_H)\sim\exp(-8\pi M/M_{\rm P})$, up to greybody and backreaction factors: astronomically small for astrophysical masses, but nonzero. This sets the correct four-dimensional baseline, since an entropy increase is only a necessary condition, whereas the physical rate of producing a near-Planckian fragment in 4D general relativity is parametrically dominated by this exponentially small Hawking/hoop probability. The beyond-GR channels are of interest precisely because they can parametrically reduce this suppression for sufficiently small $M$ or sufficiently asymmetric mass splits.}

\subsection{Bumpy horizons}\label{sec:bumpy}
We have considered so far only initial black holes that are axisymmetric. However, in the presence of non-vacuum solutions, for instance with a cosmological constant or, more generally, with exotic matter sources, bumpy horizon geometries can arise, {either as phenomenological parametrizations of deformed horizons~\cite{collins:2004ex, Johannsen:2011dh,Vigeland:2011ji} or as explicit solutions sourced by exotic matter, a cosmological constant, or higher-dimensional/higher-derivative dynamics~\cite{Novikov:1992gow, Dubovsky:2007zi, Emparan:2014pra, Suh:2023xse,Hristov:2023rel,Farhangkhah:2014zka}}. In a recent work~\cite{Canfora:2026col}, the authors have shown that such bumpy horizon configurations can also be supported by ordinary, physically realistic matter. In particular, several horizon topologies are allowed: for the purposes of the current work we restrict our attention to the spherical case. The geometry of these solutions supports a superfluid pion configuration that develops vortex structures. These vortices induce deviations from exact spherical symmetry of the horizon through their topological charges, which in turn modify the ADM mass and the entropy as follows:
\begin{equation}
    M_{\rm ADM} = \left(1- \frac{K}{4} \sum_{i \geq 1}|q_i| \right)m \qq{and} S= 4 \pi\left(1- \frac{K}{4} \sum_{i \geq 1}|q_i| \right)m^2 \ ,
\end{equation}
where $m$ is the mass parameter and $K = 4 \pi \gnewton f_\pi^2 $, with $f_\pi$ being the pion decay constant and $q_i$ the topological charges of the $i$-th defect. Additionally, for topologically spherical horizons, the constraint $\frac{K}{4} \sum_{i \geq 1}|q_i| < 1$ must hold. The formation of these vortices can lead to entropically allowed fragmentation of  the horizon. 

The easiest case to see is the splitting of a non-spinning Schwarzschild black hole with initial ADM mass $M_0$ and zero topological charges, into two bumpy black holes with mass parameters $m_1$ and $m_2$, and equal and opposite topological charges $|q|= 4 \delta v/K$, such that $\delta v < 1/2$.
By conservation of the ADM mass (neglecting eventual loss into gravitational waves), we have
\begin{equation}
    M_0 = (1-\delta v)(m_1 + m_2) \ ,
\end{equation}
which allows us to write the new black hole mass parameters in terms of the initial ADM mass, the normalized topological charge $\delta v$ and the mass ratio $r \leq 1$:
\begin{equation}
    m_1=\frac{M_0}{1-\delta v}\, r \qq{and} m_2 =\frac{M_0}{1-\delta v}(1-r) \ .
\end{equation}
The entropy change during this process is given by
\begin{equation}\label{entropybumpy}
    \frac{\Delta S}{4 \pi M_0^2} = \frac{2r(r-1)+\delta v}{1-\delta v} \ .
\end{equation}
\begin{figure}
    \centering
  
    \includegraphics[width=0.5\linewidth]{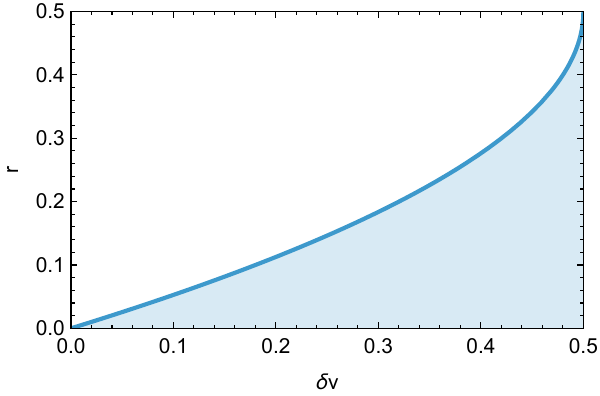}
    \caption{Region favorable to fragmentation, with increase in entropy in Eq.~\eqref{entropybumpy}, highlighted in blue in the parameter space $(r,\delta v)$. We only show $0<r<1/2$ as the region is symmetric for $r\to 1-r$, while the normalized topological charge is defined for $\delta v < 1/2$.}
    \label{bumpyhorizons}
\end{figure}
In Fig.~\ref{bumpyhorizons}, we show the parameter space in blue where the function in Eq.~\eqref{entropybumpy} is positive, as a function of the mass ratio of the black hole fragments $r$ (for $r\in[0,1/2]$ as the expression changes concavity by exchanging $r \mapsto 1-r$) and the normalized topological charge amplitude $\delta v$. We see that the condition of entropy increase $\Delta S \geq 0$ is satisfied in the region given by $2 r (1-r) \leq \delta v < 1/2$ for $0<r\leq1/2$.

\section{Conclusion}
\label{sec:conclusions}
In this work we have systematically investigated the conditions under which black hole fragmentation may occur, mapping the boundary between what is forbidden, what is merely unlikely, and what becomes possible once the standard assumptions are relaxed.
 
Within four-dimensional GR, the absence of fragmentation is robust for Schwarzschild black holes due to the Bekenstein--Hawking entropy. For Kerr black holes, entropy considerations alone favor fragmentation once the initial spin is converted into orbital angular momentum. Nevertheless, we have shown that energy and angular momentum conservation forbid the fragments from escaping to infinity, unless the parent Kerr black hole is near extremal. Instead, if produced, fragments are inevitably reabsorbed with a notable reabsorption timescale proportional to the inverse of the fragment mass: hence only fragments below the Planck mass may evaporate before reabsorption. 
 
It is well known that the situation changes dramatically in higher-dimensional GR. Black strings and ultra-spinning Myers--Perry black holes in $D>5$ are subject to the Gregory--Laflamme instability, whose onset and beads-on-a-string development are confirmed numerically. We have discussed how such processes would appear to a four-dimensional observer and shown that even bulk-only fragmentation leaves an observable gravitational imprint on the brane geometry through the induced metric.
 
We have also explored novel scenarios that may support black hole fragmentation.
Superradiant instabilities with non-axisymmetric trapping could, in principle, seed effective fragmentation in four dimensions, though significant obstacles related to wavelength constraints remain. Beyond GR, modifications to the entropy functional -- from higher-derivative gravity, generic horizon-radius deformations, or logarithmic corrections -- can render fragmentation entropically favorable for sufficiently small black holes. In the case of higher-derivative gravity, we have explicitly shown that fragmentation may occur even for black holes of astrophysical masses, in accordance with current bounds on the model parameters, as long as the fragments are much lighter. 
Finally, we have shown that bumpy horizons supported by superfluid pion configurations with non-trivial topological charges provide a novel mechanism for entropically allowed fragmentation in four dimensions, as long as the fragments carry large enough topological charges.
 
In summary, while classical four-dimensional GR enforces a strict dress code for black holes, relaxing the number of spacetime dimensions, the gravitational action, or the horizon topology can open the door to fashionable black hole pants. We should stress that entropic arguments are not enough to prove the occurrence of fragmentation, especially for static horizon configurations{: an entropy increase is a necessary but not sufficient condition, exactly as a shattered glass has higher entropy yet no dynamically favored path to shattering}. A detailed dynamical analysis or simulation would be needed to assess under which conditions black hole fragments are formed. Nevertheless, the landscape of black hole pants we charted in this manuscript has particular relevance for black holes lighter than astrophysical ones, such as primordial black holes, and non-trivial kinematic configurations, such as the occurrence of black hole mergers and scattering.
The formation of fragments can therefore provide a strong signature of beyond-GR physics, and generate novel signatures in gravitational wave observations and from the potential Hawking radiation emitted by the fragments.

\section*{Acknowledgments}
We thank Enrico Cannizzaro for useful discussions.
G.C. acknowledges partial support from the project Tremplin$@$Physique 2025, funded by the Institute of Physics (INP) of the Centre National de la Recherche Scientifique (CNRS) of France. 
VV is grateful for the hospitality of Perimeter Institute where part of this work was carried out. Research at Perimeter 
Institute is supported in part by the Government of Canada through the Department of Innovation, Science and 
Economic Development and by the Province of Ontario through the Ministry of Colleges and Universities. This work 
was supported by a grant from the Simons Foundation (1034867, Dittrich).

\bibliographystyle{JHEP}
\bibliography{biblio_v3}
%\printbibliography

\end{document}